\documentclass[journal]{IEEEtran}
\usepackage{amsmath,amsfonts}
\usepackage{algorithmic}
\usepackage{algorithm}
\usepackage{array}
\usepackage[caption=false,font=normalsize,labelfont=sf,textfont=sf]{subfig}
\usepackage{textcomp}
\usepackage{stfloats}
\usepackage{url}
\usepackage{verbatim}
\usepackage{graphicx}
\usepackage{cite}
\usepackage{xcolor}
\usepackage{booktabs}
\usepackage{multirow}
\usepackage{multicol}
\usepackage{cuted}
\usepackage{cancel}
\usepackage{hyperref}
\usepackage[subtle]{savetrees}

\begin{document}

\title{The VoxCeleb Speaker Recognition Challenge: \linebreak A Retrospective}

\author{Jaesung Huh, Joon Son Chung, Arsha Nagrani, Andrew Brown, Jee-weon Jung, \linebreak Daniel Garcia-Romero, Andrew Zisserman
\thanks{
This work is funded by the EPSRC programme grant EP/T028572/1 VisualAI and by the MSIT, Korea, under the ITRC (Information Technology Research Center) support program (IITP-2024-RS-2023-00259991) supervised by the IITP. 
For the purpose of Open Access, the author has applied a CC BY public copyright licence to any Author Accepted Manuscript version arising from this submission.
}
\thanks{Jaesung Huh and Andrew Zisserman are with Visual Geometry Group, University of Oxford, 25 Banbury Rd, Oxford, Oxfordshire, UK, OX2 6NN (e-mail: \href{mailto:jaesung@robots.ox.ac.uk}{jaesung@robots.ox.ac.uk};\href{mailto:az@robots.ox.ac.uk}{az@robots.ox.ac.uk}).}
\thanks{Arsha Nagrani was at Visual Geometry Group, University of Oxford when this work was done. She is currently with Google Research, Cambridge, MA 02142, USA (e-mail:\href{mailto:anagrani@google.com}{anagrani@google.com}).}
\thanks{Andrew Brown was at Visual Geometry Group, University of Oxford when this work was done. He is currently with GenAI, Meta, New York, NY 10003 USA (e-mail: \href{mailto:aebrown@meta.com}{aebrown@meta.com}).}
\thanks{Joon Son Chung is with the Korea Advanced Institute of Science, and
Technology, Daejeon 34141, South Korea (e-mail: \href{mailto:joonson@kaist.ac.kr}{joonson@kaist.ac.kr}).} 
\thanks{Jee-weon Jung is with Carnegie Mellon University, 5000 Forbes Ave, Pittsburgh, Pennsylvania, USA (e-mail: \href{mailto:jeeweonj@ieee.org}{jeeweonj@ieee.org}).}
\thanks{Daniel Garcia-Romero was at Johns Hopkins University, USA when this work was done. He is now with AWS AI, Arlington, VA 22202 USA (e-mail: \href{mailto:dgromero@amazon.com}{dgromero@amazon.com}). }}

\maketitle

\begin{abstract}
The VoxCeleb Speaker Recognition Challenges (VoxSRC) were a series of challenges and workshops that ran annually from 2019 to 2023.
The challenges primarily evaluated the tasks of speaker recognition and diarisation under various settings including: closed and open
training data; as well as supervised, self-supervised, and semi-supervised training for domain adaptation. 
The challenges also provided publicly available training and evaluation datasets for each task and setting, with new test sets released each year.

In this paper, we provide a review of these challenges that covers:  what they explored; the methods developed by the 
challenge participants and how these evolved; and also the current state of the field for speaker verification and diarisation. 
We chart the progress in performance over the five installments of the challenge on a common evaluation dataset and provide a detailed analysis of how each year's special focus affected participants' performance. 

This paper is aimed both at researchers who want an overview of the speaker recognition and diarisation field, and also at challenge organisers who want to benefit from the successes and avoid the mistakes of the VoxSRC challenges. 
We end with a discussion of the current strengths of the field and open challenges.
Project page : \url{https://mm.kaist.ac.kr/datasets/voxceleb/voxsrc/workshop.html}
\end{abstract}

\begin{IEEEkeywords}
Speaker verification, Speaker diarisation
\end{IEEEkeywords}

\section{Introduction}
The VoxCeleb Speaker Recognition Challenges (VoxSRC) were a series of challenges held annually from 2019 to 2023~\cite{chung2019voxsrc, nagrani2020voxsrc, brown2022voxsrc,huh2023voxsrc}. 
The goals of the challenges were threefold: (i) to explore and promote novel research in the field of speaker recognition and diarisation, encouraging important directions such as self-supervised learning and domain adaptation; (ii) to measure and calibrate the state of the art through public evaluation tools; and (iii) to provide free and open-source data to the community, accessible to all. 
The primary tasks were \textit{speaker verification} (``do these two speech segments come from the same speaker?'') and \textit{speaker diarisation} (``label a multispeaker segment with who speaks when''). 
VoxSRC consisted of an annual competition and workshop (co-located with the Interspeech conference), where each year's results and methods were discussed.   

When the challenge was first introduced in 2019, there were already a number of noteworthy speaker recognition challenges, such as those organised by the National Institute
in Standards of Technology (NIST)~\cite{alvin2004nist, greenberg2012nist, nist2018, sadjadi20172016, sadjadi20202019}, and Speakers In the Wild (SITW)~\cite{mclaren2016speakers} and speaker diarisation challenges, such as DIHARD~\cite{sell2018diarization}.
While these challenges provide immeasurable value to the community, the goal of VoxSRC was to provide complementary support in the form of open-source contributions -- all the training and validation data was (and will continue to be) free and available to
researchers irrespective of whether they enter the challenges
or not, and evaluation is performed via a public leaderboard that is visible to all. 
VoxSRC workshops were also \textit{free} for participants to attend.
The focus of VoxSRC has been on unconstrained speech from the web, with data consisting of noisy, varied and sometimes very short and fleeting speech segments. 

This paper serves as a retrospective on all five VoxSRC challenges, including their mechanics, methods, results, and discussion. 
We hope that this paper will be useful for two audiences: (i) \textit{speaker recognition researchers} aiming to determine what the state of the art in the field has been over the last five years, as measured by performance on the VoxSRC datasets, as well as wishing to understand how best practises and techniques have evolved during this period; (ii) \textit{challenge organisers}, for whom we hope our learnings serve as a useful guide for the organisation of future challenges. 
The structure of this report is as follows: we begin by describing the two primary tasks, \textit{speaker verification} and \textit{speaker diarisation} and tracks we have hosted over the last five years (Section~\ref{sec:tracks}). 
We then describe the datasets (Section~\ref{sec:datasets}), and challenge mechanics (Section~\ref{sec:mechanics}), followed by results for each track (Section~\ref{sec:results}) and a detailed analysis of the winners' method  (Section~\ref{sec:analysis}).
We describe how we hosted the workshops (Section~\ref{sec:workshop}), and finally we reflect on the trends over the years of the challenge, discuss current challenges in speaker recognition, and conclude with lessons for future challenge organisers (Section~\ref{sec:discussion}).

\begin{figure*}[!t]
    \centering
    \includegraphics[width=\textwidth]{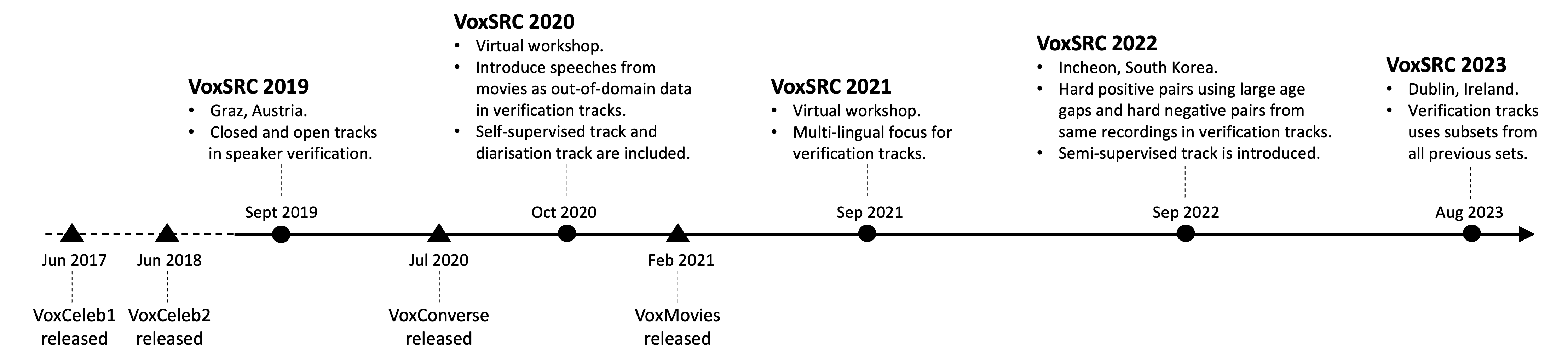}
    \caption{Timeline showing the progression of the VoxSRC workshops (dots), as well as when key datasets were released (triangles).}
    \label{fig:voxsrc_progress}
\end{figure*}

\section{Tasks and Tracks}
\label{sec:tracks}

This section provides an overview of the tracks hosted over the past five years, covering two primary challenge tasks: \textit{speaker verification} and \textit{speaker diarisation}. 
Within the speaker verification task, there were four different tracks over the last four years depending on the type of data participants are allowed to use.
In the following, we first specify the task description, and then describe the dataset constraints and the motivation behind each of the tracks.
Please also refer to Figure~\ref{fig:voxsrc_progress} which shows the progression of the VoxSRC workshops.

\subsection{Speaker verification (2019--2023)}
\label{subsec:track_verification}
The speaker verification task is to determine whether two utterances are spoken by the same speaker or not. 
The evaluation is conducted using a list of utterance pairs (i.e.\ trials), and each trial is processed independently. 
VoxSRC participants submitted a real-valued prediction score for each trial pair so that evaluation could be performed using our metrics (See Section~\ref{sec:mechanics}).
Four tracks were defined based on the choice of training data: (i) Speaker verification -- closed, (ii) Speaker verification -- open, (iii) Self-supervised speaker verification, and (iv) Semi-supervised domain adaptation.

\subsubsection{Speaker verification -- closed (2019--2023)}
\label{subsubsec:track_closed}
This has been the main challenge track since the first VoxSRC challenge in 2019.
It is closed in that for training their systems, participants were only allowed to use the publicly available VoxCeleb2~\cite{Chung18a} dev set, which spans 1,092,009 utterances from 5,994 different speakers.
The motivation here was to enable comparisons between training algorithms and model architecture approaches while keeping the data fixed.

There were two reasons for choosing only VoxCeleb2 dev set and not the rest of VoxCeleb. 
First, we used the utterances from VoxCeleb1 as the validation set, and we did not want to have any overlap between train and validation set in our setting.
Second, the size of VoxCeleb2 dev set is adequate to train large models.
The details of VoxCeleb2 are described in Section~\ref{sec:datasets}.

\subsubsection{Speaker verification -- open (2019--2023)}
\label{subsubsec:track_open}
In this track, participants were permitted to use any other data in addition to the VoxCeleb2 dev set for training \textit{except} the test data.
Regardless of whether the training data is public or not, participants here were encouraged to achieve state-of-the-art performance, pushing the limit every year.
The test data was the same as that used for the closed track to quantify the effect of external training data on speaker models.

\subsubsection{Self-supervised speaker verification (2020--2021)}
\label{subsubsec:track_selfsupervised}
In response to the progress of self-supervised learning approaches in diverse domains such as vision~\cite{chen2020simple, grill2020bootstrap, he2020momentum} and natural language processing~\cite{devlin2019bert,brown2020language}, we introduced a self-supervised speaker verification track to investigate methods for training speaker verification models without labels.
Participants in this track were only allowed to use the VoxCeleb2 dev set \textit{without} labels to train the model.
The test data was identical to the previous two tracks, which allowed the 
performance gap to be studied between methods training with and without labels.

\subsubsection{Semi-supervised domain adaptation (2022--2023)}
\label{subsubsec:track_semisupervised}
This domain adaptation track aimed to assess how models pretrained on large labelled data in a source domain can adapt to a new target domain, given (i) a large set of unlabelled data from the target domain and (ii) a small set of labelled data from the target domain. 
This setting was especially relevant to low-resource real-world scenarios, where large-scale labelled data in the source domain, and a small set of labelled data in the target domain are available in addition to large-scale unlabelled data in the target domain. 
Specifically, we focused on the task of speaker verification from one language which is the source domain (mainly English), to a different language in a target domain (Chinese).
Here we used VoxCeleb2~\cite{Chung18a} as the source domain and CNCeleb~\cite{li2022cn} as the target domain.
Participants were allowed to use the VoxCeleb2 dev set \textit{with} speaker labels, a large subset of CNCeleb \textit{without} speaker labels which contains 454,946 utterances from 1,807 speakers, and a small subset of CNCeleb \textit{with} speaker labels which consists of 1,000 utterances from 50 speakers.

\subsection{Speaker diarisation (2020--2023)}
The speaker diarisation task is to identify ``who speaks when'' given audio containing single or multi-speaker speech segments.
Participants were required to (i) identify speaker regions, and (ii) assign a speaker label to each region in the audio file.
Participants did not need to match speech segments to specific known speakers; instead, they could simply cluster the segments according to different speakers.
Participants were allowed to use any data to train their models \textit{except} the test data.

\section{Datasets}
\label{sec:datasets}
This section describes the training, validation and test sets used in the challenges.
Each year, we created (i) validation sets for which labels were available to the public, and (ii) test sets constructed from hidden data to make the task challenging and interesting, for which labels were not available publicly.
For each track, the validation set featured a data distribution similar to that of the corresponding test set.
This section details the datasets used each year, encompassing both publicly available and organiser-created hidden datasets. 
Additionally, we include a discussion of the annotation methods.
We show the training set for each track in Table~\ref{tab:track_trainset}. 
The complete statistics of the validation and test sets each year are given in the Appendix~\ref{appendix:stats_valtest}.

During the VoxSRC challenges, the labels of the test sets were not released. However, following the end of the challenges, they have now been released for the community.\footnote{\url{https://mm.kaist.ac.kr/datasets/voxceleb/voxsrc/workshop.html}}

\begin{table}[t]
\caption{Training sets per each track. \linebreak Statistics of validation and test sets are in the Appendix~\ref{appendix:stats_valtest}.}
\label{tab:track_trainset}
\centering
\resizebox{1\linewidth}{!}{
\begin{tabular}{@{}lll@{}}
\toprule
\textbf{Track}                                & \textbf{Year}                 & \textbf{Training set}                                                                                                                                   \\
\midrule
Speaker verification -- closed        & 2019 -- 2023          & VoxCeleb2 dev set                                                                                                                           \\ \midrule
Speaker verification -- open          & 2019 -- 2023 & Any data except the test audio files                                                                                                              \\ \midrule
Self-supervised speaker verification & 2020 -- 2021 & VoxCeleb2 dev set without labels                                                                                                            \\ \midrule
Semi-supervised domain adaptation    & 2022 -- 2023          & \begin{tabular}[c]{@{}l@{}}- VoxCeleb2 dev set with labels\\ - A large subset of CNCeleb2 without labels\\ - A small subset of CNCeleb1 with labels\end{tabular} \\
\midrule
Speaker diarisation                  & 2020 -- 2023          & Any data except the test audio files \\ 
\bottomrule
\end{tabular}}
\end{table}

\subsection{Speaker verification (Tracks 1 and 2)}
\label{subsec:dataset_veri}
The challenge datasets were based on VoxCeleb~\cite{Nagrani17,Chung18a,nagrani2020voxceleb}, a large-scale speaker recognition dataset comprising utterances from celebrities, sampled from interviews and TV shows on YouTube. 
The dataset was created by extracting multiple single-speaker utterances from each YouTube video using face tracking~\cite{Liu16}, face verification~\cite{Cao18} and active speaker detection~\cite{Chung16a}.
This dataset consists of two versions. 
VoxCeleb1~\cite{Nagrani17} includes over 150,000 utterances from 1,251 speakers in 22,496 unique recordings. 
The audio dataset covers diverse background environments and recording conditions.
VoxCeleb2~\cite{Chung18a} expands the initial version to a larger scale, containing over a million utterances from 6,112 speakers in 150,480 recordings, making it five times larger than VoxCeleb1.

\subsubsection{Training set}
For the closed track (Track 1), participants were only allowed to use the VoxCeleb2 dev set for training their systems.\footnote{Note, the ``VoxCeleb2 dev set'' used in VoxSRC refers to the training set of the original VoxCeleb2 dataset.}
The dev set contains 1,092,009 utterances from 5,994 speakers. 
For the open track (Track 2), participants could use any data except the challenge test set.
The training set on these two tracks remained the same for all five years.

\subsubsection{Validation set} 
The validation sets for all five years were based on VoxCeleb1. 
In addition, the validation set for 2020 and 2023 challenges made use of audio clips from the VoxMovies dataset, and the 2022 and 2023 challenges utilised audio segments from the VoxConverse dataset.

\subsubsection{Test set}
The test data was created from YouTube videos in the same way as the training and validation sets, but they came from identities that do not appear in VoxCeleb1 or VoxCeleb2. 
The test data was checked manually for any errors using the same procedure described in~\cite{Nagrani17}. 
This was done using a simple web-based tool that shows all video segments for each identity. 
To monitor the performance improvements over time, the VoxSRC 2019 test trials were always included in the test sets of all subsequent years.

\subsubsection{Themes}
Each year's challenge was designed to emphasise important research directions. 
The following paragraphs provide additional information on how the data was collected for each challenge.

\noindent\textbf{Speakers at the Movies (2020).}
In VoxSRC 2020, we introduced a new set of speaker segments from movie material, namely the VoxMovies dataset~\cite{brown2020playing}, into the validation and test sets, which served as out-of-domain data for some of the identities in VoxCeleb. 
VoxMovies is a speaker recognition dataset comprising utterances from movies, collected using a similar pipeline to VoxCeleb.
It poses greater challenges than VoxCeleb because actors tend to disguise their voices and put more emotions in their speech in movies compared to in interviews.
The utterances vary in emotion, accent, and background noise and therefore come from an entirely different domain to the VoxCeleb utterances, which mostly contain celebrity interviews.

Utterances trials were constructed from the hidden sets of VoxMovies and VoxCeleb to create more challenging test pairs. 
Pairs were constructed in two different ways: (i) both utterances originated from either VoxCeleb or VoxMovies; or (ii) one utterance from VoxCeleb and the other from VoxMovies.

\noindent\textbf{Cross-lingual pairs (2021).}
In VoxSRC 2021, multi-lingual verification pairs were introduced into the speaker verification validation and test data.
The goals of this were twofold: first to promote the fairness and accessibility of speaker verification models, so as to allow people from diverse language groups to use these deep learning models; and second, to provide a more challenging test set for the speaker verification tracks.

Due to the design of the dataset collection pipeline~\cite{nagrani2020voxceleb}, the VoxCeleb datasets consist of mainly English-speaking speech segments. 
In the validation and test sets in 2021, a multi-lingual focus was added by sampling more positive and negative pairs that contain non-English speech segments. 
This required the use of language labels, which do not exist for the VoxCeleb datasets. 
We obtained the language labels using a three-step pipeline consisting of a combination of automatic and manual annotation. 
First, we obtained automatic language predictions for each VoxCeleb video using a model trained on VoxLingua107~\cite{valk2021slt} to make language predictions across 107 languages. 
We assumed each speaker in a video uses only one language and randomly selected one utterance per video together with its predicted language label. 
Second, we manually annotated the correctness of the language predictions for the 12 most frequently occurring languages using 
a customised LISA annotation tool~\cite{duta20lisa}.
The 12 languages include English, French, Dutch, Italian, German, Spanish, Hindi, Portuguese, Russian, Chinese, Japanese, and Korean.
Third, we used these manual annotations to obtain language-specific classification thresholds for the automatic predictions. 
These thresholds were then applied to classify each speech segment in VoxCeleb1 as either one or none of these 12 languages.

\noindent\textbf{Hard positive pairs with large age gaps (2022). }
We curated hard positive validation and test pairs where the age of the speaker differs considerably between the two utterances. 
These hard positives were found by selecting utterance pairs from the same speaker that have a large age gap (i.e.\ two audio files for the same identity where the age is very different) via a two-step process in the hidden video data. 
In this data, for each video segment, we had the face location, the identity and the audio recording.
First, the age of the speaker was estimated by predicting the age for a random set of frames, using an open-source face age prediction network~\cite{FaceLibGit}, and averaging the result. 
Second, we sampled positive pairs from utterances for the same speaker with large age gaps.
We performed this process within VoxCeleb1 to create the validation set, and within the hidden data to create the test set.

\noindent\textbf{Hard negative pairs that share the same background noise (2022).}
We constructed hard negative validation and test pairs by sampling utterances from different speakers within the same video that therefore share very similar acoustic conditions.
Most of the negative pairs in the VoxCeleb training datasets are from different videos, therefore speaker verification systems might be able to use the environmental cues as a shortcut for speaker prediction.
Our goal here was to construct harder negative pairs by sampling utterances from different speakers that are sourced from the same video. 
In this case, the environmental conditions are shared across the two utterances and only the speaker's identity changes. 
We sampled the hard negative pairs using the VoxConverse~\cite{Chung20} speaker diarisation dataset, where each audio file consists of multiple short speech segments from different speakers~\cite{jung2023search}.
To generate these trials, we first cropped short speech segments.
We then removed segments that were either too short ($<$1.5s) or had overlapping speech.
Finally, we selected trials using two segments from different speakers within the same audio file.

\noindent\textbf{Combination of all themes (2023).}
In the final year, we composed a test set including all scenarios aforementioned, making it the most comprehensive test set.
This includes all test pairs from the years 2019, 2021, 2022, and 2023, as well as approximately 200,000 randomly selected pairs from 2020. 
To avoid redundancy, we ensured that each pair in the VoxSRC 2019 test set was included only once.
Note, only 200,000 of the approximately 1.7 million pairs in the VoxSRC2020 test set are included in order to balance the number of test pairs for each year, and also to reduce the computational expense of evaluation.

\subsection{Self-supervised speaker verification (Track 3 in 2020 and 2021)}

For the self-supervised track introduced in 2020, participants were allowed to use VoxCeleb2 dev set \textit{without} labels. 
Training, validation and test sets are otherwise identical to Track 1 and 2 in the respective years.

\subsection{Semi-supervised domain adaptation (Track 3 in 2022 and 2023)}

As described in Section~\ref{subsubsec:track_semisupervised}, the domain adaptation that we focused on was from one language in the source domain (mainly English) to a different language in the target domain (Chinese).
For this challenge, the VoxCeleb dataset represents the source domain, and
CNCeleb~\cite{li2022cn} represents the target domain.
CNCeleb is a large-scale speaker verification dataset, mostly from Chinese speakers, which contains more than 600,000 utterances from 3,000 identities.
Among the 11 genres CNCeleb spans,  we removed ``singing'', ``play'', ``movie'', ``advertisement'' and ``drama'' genres to focus on language domain adaptation tasks. 

\subsubsection{Training set}
Participants were allowed to use three types of datasets in this track:

\begin{itemize}
  \item VoxCeleb2 dev set \textit{with} speaker labels (source domain). This can be used for pretraining.
  \item A large subset of CNCeleb2 \textit{without} speaker labels (target domain). 
  It consists of 454,946 utterances from 1,807 identities. This can be used for domain adaptation.
  \item A small subset of CNCeleb1 \textit{with} speaker labels (target domain) comprising 50 speakers with 20 utterances per speaker.
\end{itemize}

\subsubsection{Validation set}
We provided a list of trial speech pairs from identities in the target domain. 
These utterances were sampled from CNCeleb1.

\subsubsection{Test set}
The test set consists of 56 disjoint identities not present in either CNCeleb1 or CNCeleb2.
We used the hidden set of CNCeleb, provided by the authors, and constructed 30,000 and 80,000 pairs in years 2022 and 2023, respectively.

\subsection{Speaker diarisation (Track 4 in 2020 to 2023)}

The VoxConverse~\cite{Chung20} dataset is an audio-visual speaker diarisation dataset which includes 448 videos from YouTube.
These videos are mostly from debates, talk shows and news segments. 
It has multi-speaker, variable-length audio segments, with overlaps and challenging acoustic conditions. 
Inspired by other audio-visual dataset creation pipelines such as VoxCeleb~\cite{nagrani2020voxceleb} and VGGSound~\cite{chen2020vggsound}, the pipeline leverages an automatic audio-visual speaker diarisation method using active speaker detection~\cite{Chung16a}, audio-visual source separation~\cite{Afouras18} and speaker verification~\cite{chung2020defence}, followed by manual verification.
Only audio files were provided for this challenge. 

\subsubsection{Training set} Participants were allowed to use any public or internal datasets \textit{except} for the test data to train their systems.

\subsubsection{Validation set} We provided the dev set of VoxConverse as the validation set for the 2020 challenge.
The participants were allowed to use the entire VoxConverse as the validation set for the remaining years.

\subsubsection{Test set} 
In the year 2020, we used the VoxConverse test set as our challenge test set, which was included in the validation set for subsequent years. 
The test sets for the following years were curated using the same pipeline as VoxConverse, but we utilised a hidden set of videos that were not  public. 
These videos, sourced from YouTube, span diverse categories including news, documentaries, lectures, and commercials. 
We curated test sets with 264, 360, and 413 audio files in the years 2021, 2022, and 2023, respectively.

\section{Challenge mechanics}
\label{sec:mechanics}
This section outlines the challenge evaluation metrics for all tracks, followed by an overview of how we hosted the challenge including submission rules and formats.

\subsection{Evaluation metrics}
Each year, we released a validation toolkit to allow participants to assess their systems.
The code in this toolkit was identical to the one which organisers use for evaluation, preventing possible performance differences stemming from implementation mismatches. 
All verification tracks shared the same metrics, minimum Detection Cost Function (minDCF) and Equal Error Rate (EER).
The diarisation track used Diarisation Error Rate (DER) and Jaccard Error Rate (JER) as evaluation metrics. The evaluation protocols and metrics are adopted, with minor modifications, directly from NIST SRE challenges~\cite{nist2018} for the verification tracks, and from DIHARD~\cite{sell2018diarization} for the diarisation track.

\subsubsection{Speaker verification}
\textbf{minDCF} is the calibration insensitive metric to measure speaker verification performance.
The DCF is computed as:
\begin{equation}
    C_{DET} = C_{miss} \times P_{miss} \times P_{tar} +C_{fa} \times P_{fa} \times (1 - P_{tar}) 
\label{eqn:dcf}
\end{equation}

$P_{miss}$ and $P_{fa}$ are normalised error rates by counting the errors from positive and negative pair trials respectively, and $P_{tar}$ is a prior probability that a target speaker event occurs in the real world, which is provided by the evaluator.
minDCF is the minimum value of $C_{DET}$ by varying the threshold.
We set $C_{miss} = C_{fa} = 1$ and $P_{tar}=0.05$ in our cost function.
The value of $P_{tar}=0.05$ is similar to the proportion of positive pairs (4\%) in the VoxSRC 2019 test set, and is also used in NIST SRE evaluation.
\textbf{EER} corresponds to the value where False Acceptance (FA) and False Rejection (FR) rates are equal. 
EER is independent of parameters, unlike minDCF which is dependent on a set of predefined parameters such as $C_{miss}$,$C_{fa}$ and $P_{tar}$.

\subsubsection{Speaker diarisation}
\textbf{DER} is the standard metric for evaluating diarisation results between prediction (i.e. hypothesis) and ground truth (i.e. reference). 

It is computed as:

\begin{equation}
    DER = \frac{\texttt{MISS + FA + CONF}}{\texttt{Reference}}
\label{eqn:der}
\end{equation}

where \texttt{Reference} is the total length of reference, \texttt{MISS} is the total length of speech that is present in reference but not in hypothesis and \texttt{FA}, false alarm, is the total length of speech that is present in hypothesis but not in reference.
\texttt{CONF}, denoting confusion, refers to the total duration of speech which the system incorrectly identifies the speakers.
We applied a forgiveness collar of 0.25 seconds when computing the DER to compensate for small inconsistencies in the annotation.
Overlapping speech was considered during the DER calculation.

\textbf{JER} was newly introduced in the DIHARD II challenge~\cite{ryant2019second} as another diarisation metric.
It is adopted from the Jaccard similarity index, which is used to evaluate image segmentation.
To compute this, we first construct the mapping between speakers in reference and hypothesis using the Hungarian algorithm.
Then for each reference speaker, we compute $JER_{spk}$ using Equation~\ref{eqn:jer} where \texttt{Total} is the duration of the union of speaker segments in reference and hypothesis, \texttt{MISS} is the duration of speaker segments in the reference which are not present in hypothesis, and \texttt{FA} is the duration of speaker segments in the hypothesis which do not exist in reference.
The total JER is the average of $JER_{spk}$.

\begin{equation}
    JER_{spk} = \frac{\texttt{MISS + FA}}{\texttt{Total}}
\label{eqn:jer}
\end{equation}

\subsection{How we hosted the VoxSRC challenge}
The challenge started approximately two months before the workshop when the validation and test sets for each track were released.
The submission format differed between speaker verification and diarisation tracks.
For speaker verification tracks, we provided a list of pairs with corresponding audio files and asked the participants to submit the prediction scores for whether each pair was from the same speaker or not.
For speaker diarisation tracks, participants were required to submit their results in RTTM format, which contains channel id, start time, duration and speaker id for each speech segment.
We also provided a validation set per track with ground truth labels to let participants measure their systems' performance during development.

The challenge was hosted via CodaLab~\cite{codalab_competitions}.
Each year we used the submission server provided by the CodaLab platform except for the last two years when we created our own backend to deal with bugs and errors effectively.

Our challenge had two phases: the \textit{challenge} phase and the \textit{permanent} phase.
The \textit{challenge} phase ended before the workshop and only submissions made within this phase were considered for the prizes and certificates at the following workshop.
We also opened a \textit{permanent} phase after the workshop ended to let people both in industry and academia evaluate their systems with our test set and compare them against competition winners' performance.

Participants were allowed to submit only one submission per day. 
The total number of submissions allowed was 5 in 2019 and 2020, and then was increased to 10 for future challenges reflecting participants' feedback.
The limit on submission was to prevent overfitting on the test set. 
To prevent the same team from making submissions across multiple CodaLab accounts, we only allowed participants who registered with either academic or industry email accounts to participate. We show the participant statistics each year in the Appendix~\ref{appendix:participant_statistics}.

We also provided baselines with code and models for all tracks to help new participants get started each year.
We mostly used ResNet-34 models adopted from \cite{kwon2021ins} for speaker verification.
For diarisation, we adopted the clustering-based approach from \cite{lin2019lstm} using pyannote~\cite{bredin2020pyannote} VAD, speaker embedding extractor from either \cite{chung2020defence} or \cite{desplanques2020ecapa} and agglomerative hierachical clustering. 
We released both verification~\cite{voxceleb_trainer} and diarisation baseline~\cite{simplediarization} in public GitHub repositories.
\section{Results}
\label{sec:results}

\begin{table*}[!ht]
\caption{An overview of the winning methods on Tracks 1 and 2 across the VoxSRC challenges.}
\label{tab:veri_winner_method}
\resizebox{\textwidth}{!}{%
\begin{tabular}{@{}ccllllll@{}}
\toprule
\textbf{Workshop}                                                               & \textbf{Track} & \multicolumn{1}{c}{\textbf{Input feature}}                   & \multicolumn{1}{c}{\textbf{Embedding network architecture}}                                                           & \multicolumn{1}{c}{\textbf{Data augmentation}}                                                                                                                    & \multicolumn{1}{c}{\textbf{Loss function}}                                              & \multicolumn{1}{c}{\textbf{Back-end}}                                     & \multicolumn{1}{c}{\textbf{External data}}                                                 \\ \midrule
\multirow{2}{*}{\textbf{\begin{tabular}[c]{@{}c@{}}VoxSRC\\ 2019\end{tabular}}} & \textbf{1}     & \begin{tabular}[c]{@{}l@{}}PLP\\ log-mel Fbank\end{tabular}  & \begin{tabular}[c]{@{}l@{}}- ResNet~\cite{he2016deep} variants\\ - TDNN\end{tabular}                                           & \begin{tabular}[c]{@{}l@{}}- Reverberation\\ - Additive noise with MUSAN~\cite{snyder2015musan}\end{tabular}                                                                    & - AAM loss~\cite{deng2019arcface}                                                                     & \begin{tabular}[c]{@{}l@{}}- GPLDA~\cite{garcia2012multicondition}\\ - AS-norm~\cite{matejka2017analysis}\end{tabular}        & -                                                                                 \\ \cmidrule(l){2-8} 
            & \textbf{2}     & \begin{tabular}[c]{@{}l@{}}PLP\\ log-mel Fbank\end{tabular}  & \begin{tabular}[c]{@{}l@{}}- ResNet~\cite{he2016deep} variants\\ - TDNN\end{tabular}                                           & \begin{tabular}[c]{@{}l@{}}- Reverberation\\ - Additive noise with MUSAN~\cite{snyder2015musan}\end{tabular}                                                                    & - AAM loss~\cite{deng2019arcface}                                                                     & \begin{tabular}[c]{@{}l@{}}- GPLDA~\cite{garcia2012multicondition}\\ - AS-norm~\cite{matejka2017analysis}\end{tabular}        & \begin{tabular}[c]{@{}l@{}}- Librispeech~\cite{panayotov2015librispeech} \\ - Deepmine~\cite{zeinali18b_odyssey}\end{tabular}               \\ \midrule
\multirow{2}{*}{\textbf{\begin{tabular}[c]{@{}c@{}}VoxSRC\\ 2020\end{tabular}}} & \textbf{1}     & \begin{tabular}[c]{@{}l@{}}MFCC\\ log-mel Fbank\end{tabular} & \begin{tabular}[c]{@{}l@{}}- ECAPA-TDNN~\cite{desplanques2020ecapa}\\ - SE-ResNet34 variants\end{tabular}                                & \begin{tabular}[c]{@{}l@{}}- Reverberation\\ - Additive noise with MUSAN~\cite{snyder2015musan}\\ - SpecAugment~\cite{park2019specaugment}\\ - Tempo up and down\\ - Compression using FFmpeg\end{tabular} & \begin{tabular}[c]{@{}l@{}}- AAM loss~\cite{deng2019arcface}\\ - Large-margin Finetuning\end{tabular} & \begin{tabular}[c]{@{}l@{}}- QMF\\ - AS-norm~\cite{matejka2017analysis}\end{tabular} & -                                                                                 \\ \cmidrule(l){2-8} 
    & \textbf{2}     & \begin{tabular}[c]{@{}l@{}}MFCC\\ log-mel Fbank\end{tabular} & \begin{tabular}[c]{@{}l@{}}- ECAPA-TDNN~\cite{desplanques2020ecapa}\\ - Se-ResNet34 variants\end{tabular}                                & \begin{tabular}[c]{@{}l@{}}- Reverberation\\ - Additive noise with MUSAN~\cite{snyder2015musan}\\ - SpecAugment~\cite{park2019specaugment}\\ - Tempo up and down\\ - Compression using FFmpeg\end{tabular} & \begin{tabular}[c]{@{}l@{}}- AAM loss~\cite{deng2019arcface}\\ - Large-margin Finetuning\end{tabular} & \begin{tabular}[c]{@{}l@{}}- QMF\\ - AS-norm~\cite{matejka2017analysis}\end{tabular} & \begin{tabular}[c]{@{}l@{}}- VoxCeleb1~\cite{Nagrani17}\\ - Librispeech~\cite{panayotov2015librispeech} \\ - Deepmine~\cite{zeinali18b_odyssey}\end{tabular} \\ \midrule
\textbf{\begin{tabular}[c]{@{}c@{}}VoxSRC\\ 2021\end{tabular}}                  & \textbf{1,2}   & log-mel Fbank                                                & \begin{tabular}[c]{@{}l@{}}- RepVGG~\cite{ding2021repvgg}\\ - ResNet~\cite{he2016deep} variants\end{tabular}                                         & \begin{tabular}[c]{@{}l@{}}- Reverberation\\ - Additive noise with MUSAN~\cite{snyder2015musan}\\ - White noise\\ - Gain augment\\ - Time stretch augment\end{tabular}          & \begin{tabular}[c]{@{}l@{}}- AAM loss~\cite{deng2019arcface}\\ - Inter-TopK loss\end{tabular}         & \begin{tabular}[c]{@{}l@{}}- QMF\\ - AS-norm~\cite{matejka2017analysis}\end{tabular} & -                                                                                 \\ \midrule
\multirow{2}{*}{\textbf{\begin{tabular}[c]{@{}c@{}}VoxSRC\\ 2022\end{tabular}}} & \textbf{1}     & log-mel Fbank                                                & - ResNet~\cite{he2016deep} variants                                                                                            & \begin{tabular}[c]{@{}l@{}}- Reverberation\\ - Additive noise with MUSAN~\cite{snyder2015musan}\\ - Speed augment\\ - SpecAugment~\cite{park2019specaugment}\end{tabular}                                  & - AM loss~\cite{wang2018additive}                                                                        & \begin{tabular}[c]{@{}l@{}}- QMF\\ - AS-norm~\cite{matejka2017analysis}\end{tabular} & -                                                                                 \\ \cmidrule(l){2-8} 
    & \textbf{2}     & log-mel Fbank                                                & \begin{tabular}[c]{@{}l@{}}- ResNet~\cite{he2016deep} variants\\ - ECAPA-TDNN~\cite{desplanques2020ecapa} with \\   WavLM~\cite{chen2022wavlm} and Hubert~\cite{hsu2021hubert} features\end{tabular} & \begin{tabular}[c]{@{}l@{}}- Reverberation\\ - Additive noise with MUSAN~\cite{snyder2015musan}\\ - Speed augment\\ - SpecAugment~\cite{park2019specaugment}\end{tabular}                                  & \begin{tabular}[c]{@{}l@{}}- AM loss~\cite{wang2018additive}\\ - AAM loss~\cite{deng2019arcface}\end{tabular}                 & \begin{tabular}[c]{@{}l@{}}- QMF\\ - AS-norm~\cite{matejka2017analysis}\end{tabular} & Self-VoxCeleb                                                                     \\ \midrule
    \multirow{2}{*}{\textbf{\begin{tabular}[c]{@{}c@{}}VoxSRC\\ 2023\end{tabular}}} & \textbf{1}     & log-mel Fbank                                                & \begin{tabular}[c]{@{}l@{}}- RepVGG~\cite{ding2021repvgg}\\ - ResNet~\cite{he2016deep} variants\\- Multi-query multi-head attention (MQMHA)\end{tabular} & \begin{tabular}[c]{@{}l@{}}- Reverberation\\ - Additive noise with MUSAN~\cite{snyder2015musan}\\ - Speed augment\end{tabular}                                  & \begin{tabular}[c]{@{}l@{}}- AM loss~\cite{wang2018additive}\\ - AAM loss~\cite{deng2019arcface}\end{tabular}                                                                        & \begin{tabular}[c]{@{}l@{}}- QMF\\ - AS-norm~\cite{matejka2017analysis}\end{tabular} & -                                                                                 \\ \cmidrule(l){2-8} 
    & \textbf{2}     & log-mel Fbank                                                & \begin{tabular}[c]{@{}l@{}}- ResNet~\cite{he2016deep} variants\\ - ECAPA-TDNN~\cite{desplanques2020ecapa} with   WavLM~\cite{chen2022wavlm}, \\Unispeech~\cite{chen2022unispeech} and XLSR~\cite{babu2021xls} features\end{tabular} & \begin{tabular}[c]{@{}l@{}}- Reverberation\\ - Additive noise with MUSAN~\cite{snyder2015musan}\\ - SpecAugment~\cite{park2019specaugment}\end{tabular}                                  & \begin{tabular}[c]{@{}l@{}}- AM loss~\cite{wang2018additive}\\ - AAM loss~\cite{deng2019arcface}\end{tabular}                 & \begin{tabular}[c]{@{}l@{}}- CMF score\\- QMF\\ - AS-norm~\cite{matejka2017analysis}\end{tabular} & VoxTube~\cite{yakovlev23_voxtube}\\ \bottomrule
\end{tabular}}
\end{table*}
This section discusses the results of each track over the last five years.
We also study the longitudinal progress on verification and diarisation tracks over the years by comparing performance on the VoxSRC 2019 verification\footnote{Included as a subset in all subsequent years' test sets.} and 2021 diarisation test sets.

\subsection{Speaker verification -- Track 1 and 2}
\label{subsec:result_track12}

Table~\ref{tab:veri_winner_method} gives an overview of the winners' system for Tracks 1 and 2, comparing them over six aspects. The performance of winners each year is shown in Table~\ref{tab:results_comparison_verification}.
As the input, speech waveforms could be directly utilised or could be converted into acoustic features. 
All teams adopted a DNN-based embedding extractor, which maps a variable length speech segment into a single speaker representation.
Participants trained the DNN with diverse architectures such as ResNet~\cite{he2016deep}, ECAPA-TDNN~\cite{desplanques2020ecapa} or RepVGG~\cite{ding2021repvgg} and combine several model outputs for final submission.
Commonly leveraged data augmentation techniques included additive noise with MUSAN~\cite{snyder2015musan}, reverberation using public room impulse response (RIR) dataset~\cite{ko2017study} or spec augmentation~\cite{park2019specaugment}.

Angular Margin softmax loss (AM loss)~\cite{wang2018additive} and Additive Angular Margin softmax loss (AAM loss)~\cite{deng2019arcface} were the two most commonly used training objectives to learn the embedding extractor.
A few winners adopted additional techniques such as large-margin finetuning~\cite{thienpondt2020idlab} or Inter-TopK penalty~\cite{zhao2021speakin} to further improve their model performance.

The scoring procedures included probabilistic linear discriminant analysis (PLDA), and most commonly, direct cosine scoring. Both of them optionally followed by score normalisation.
From the year 2020, the VoxSRC winners started to use the Quality Measure Function (QMF), which has already been explored in the NIST SREs~\cite{hasan2013crss, mandasari2013quality}.
They include quality metrics such as speech duration or magnitude of non-normalised embeddings to model various conditions of the trial utterances using logistic regression. 

In Track 2, several participants utilised external public data~\cite{panayotov2015librispeech, zeinali18b_odyssey}.
However, it did not show a clear performance gap between Tracks 1 and 2 until VoxSRC 2021.
Both the VoxSRC 2022 winner~\cite{ravana2022} and 2023 winner~\cite{torgashov2023id} demonstrated that self-supervised pretrained models generalise well in the new domain with a clear gap between Tracks 1 and 2 submissions (EER 1.49\% to 1.21\%).
They leveraged self-supervised pretrained models~\cite{hsu2021hubert,chen2022wavlm, chen2022unispeech, babu2021xls} to extract features and train another speaker model using these features.
The VoxSRC 2022 and 2023 winner also curated their own datasets, Self-VoxCeleb and VoxTube~\cite{yakovlev23_voxtube} respectively, using a similar data creation pipeline to that of VoxCeleb.

 \begin{table}[!t]
     \caption{Comparison of methods on the four workshop test sets of Track 1 and 2. We compare top-2 submissions of the last four years on the 2019 test set. For \% EER shown, lower is better. Note that in most years the EER is not the primary evaluation metric. Therefore, in some years the 2nd place have a better EER than the 1st place.}
     \vspace{-5pt}
    \label{tab:results_comparison_verification}
    \centering
    \resizebox{1\linewidth}{!}{
    \begin{tabular}{lcccccc}
    \toprule
         \textbf{Team} & \textbf{Track} & \textbf{2019 test} & \textbf{2020 test}   & \textbf{2021 test} & \textbf{2022 test} & \textbf{2023 test}\\ \midrule
         Baseline~\cite{kwon2021ins} & - & 1.29 & 5.01 &5.00 & 5.26 & 5.45 \\ \midrule
         VoxSRC 2019 1st place~\cite{zeinali2019but} & 1 & 1.42 & - & - & - & -\\
         VoxSRC 2019 2nd place~\cite{romero2020jhu} & 1 & 1.54 & - & - & - & -\\
         \midrule
        VoxSRC 2019 1st place~\cite{zeinali2019but} & 2 & 1.26 & - & - & - & -\\
         VoxSRC 2019 2nd place~\cite{tz2019} & 2 & 1.49 & - & - & - & -\\
         \midrule
          VoxSRC 2020 1st place~\cite{thienpondt2020idlab} & 1 & 0.83 &  3.73 & - & - & -\\  
          VoxSRC 2020 2nd place~\cite{xiang2020xx205} & 1 & 0.75 & 3.81 & - & - & -\\
         \midrule
            VoxSRC 2020 1st place~\cite{thienpondt2020idlab} & 2 & 0.80 &  \textbf{3.58} & - & - & -\\  
          VoxSRC 2020 2nd place~\cite{xiang2020xx205} & 2 & 0.79 & 3.80 & - & - & -\\
         \midrule
         VoxSRC 2021 1st place~\cite{zhao2021speakin} & 1,2 & 0.57 & - & \textbf{1.85} & - & -\\  
          VoxSRC 2021 2nd place~\cite{zhang2021beijing} & 1,2 & 0.62 & - & 2.84 & - & -\\
        \midrule
         VoxSRC 2022 1st place~\cite{ravana2022} & 1 & 0.90 & - & - & 1.49 & -\\  
          VoxSRC 2022 2nd place~\cite{cai2022kriston} & 1 & 0.65 & - & - & 1.40 & -\\
        \midrule
          VoxSRC 2022 1st place~\cite{ravana2022} & 2 & 0.69 & - & - & 1.21 & -\\  
          VoxSRC 2022 2nd place~\cite{cai2022kriston} & 2 & 0.50 & - & - & \textbf{1.12} & -\\
          \midrule
          VoxSRC 2023 1st place~\cite{zheng2023unisound} & 1 & 0.58 & - & - & - & 1.59\\  
          VoxSRC 2023 2nd place~\cite{bilibili2023} & 1 & 0.59 & - & - & - & 1.76 \\
        \midrule
          VoxSRC 2023 1st place~\cite{torgashov2023id} & 2 & \textbf{0.47} & - & - & - & \textbf{1.30} \\  
          VoxSRC 2023 2nd place~\cite{zheng2023unisound} & 2 & 0.58 & - & - & - & 1.59 \\
          \bottomrule
    \end{tabular}}
\end{table} 

\noindent\textbf{Performance progression.} The VoxSRC 2019 test set has been included as a subset in all challenges.
We analyse the progress of state of the art by comparing the winning teams' performances on this set. 

\begin{figure}[!t]
    \centering
    \includegraphics[width=\linewidth]{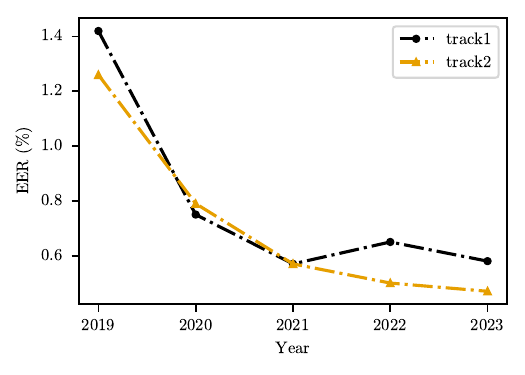}
    \caption{Top winner's performance on VoxSRC 2019 test set. We report the 2nd place's performance for VoxSRC 2020 and 2022, since they are better than the winner on VoxSRC 2019 test set. All other entries are from the 1st place.}
    \label{fig:veri_progress}
\end{figure}

Results are shown in Figure~\ref{fig:veri_progress}.
By comparing the Track~1 winning methods over the last four years, we see that state-of-the-art performance each year has steadily improved, except for the last two years.
However, when comparing Track~2 submissions, the performance has improved significantly even in the last year due to the use of public self-supervised pretrained models.

\begin{figure}[t]
    \centering
    \includegraphics[width=1\linewidth]{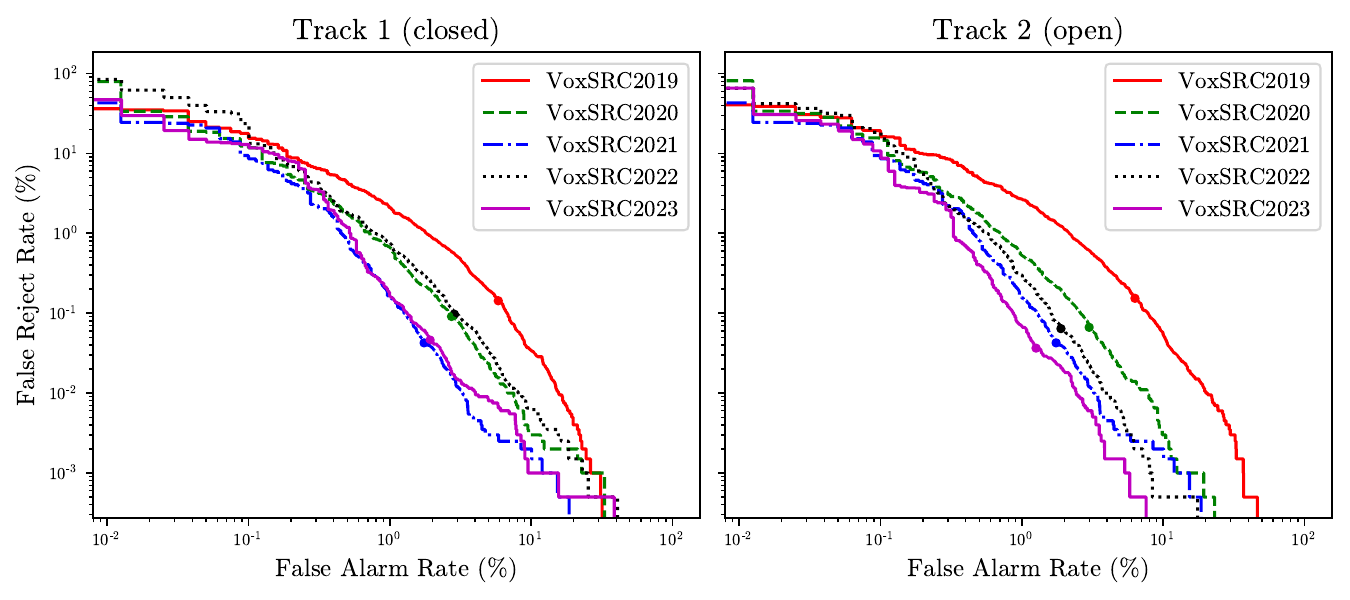}
    \caption{DET curves of the top winners each year on the VoxSRC 2019 test set. The circles are the points where the DCF value is minimum.}
    \label{fig:det_curve}
\end{figure}

Figure~\ref{fig:det_curve} shows DET curves from the winners' performance on the VoxSRC 2019 test set. The closer the curve is to the origin, the better the performance. Note the VoxSRC2019 winners perform worst on Track 1 and Track 2. The VoxSRC2021 winner has the best performance on Track 1, while the VoxSRC2023 winner is the best on Track 2.

\begin{figure}[t]
    \centering
    \includegraphics[width=\linewidth]{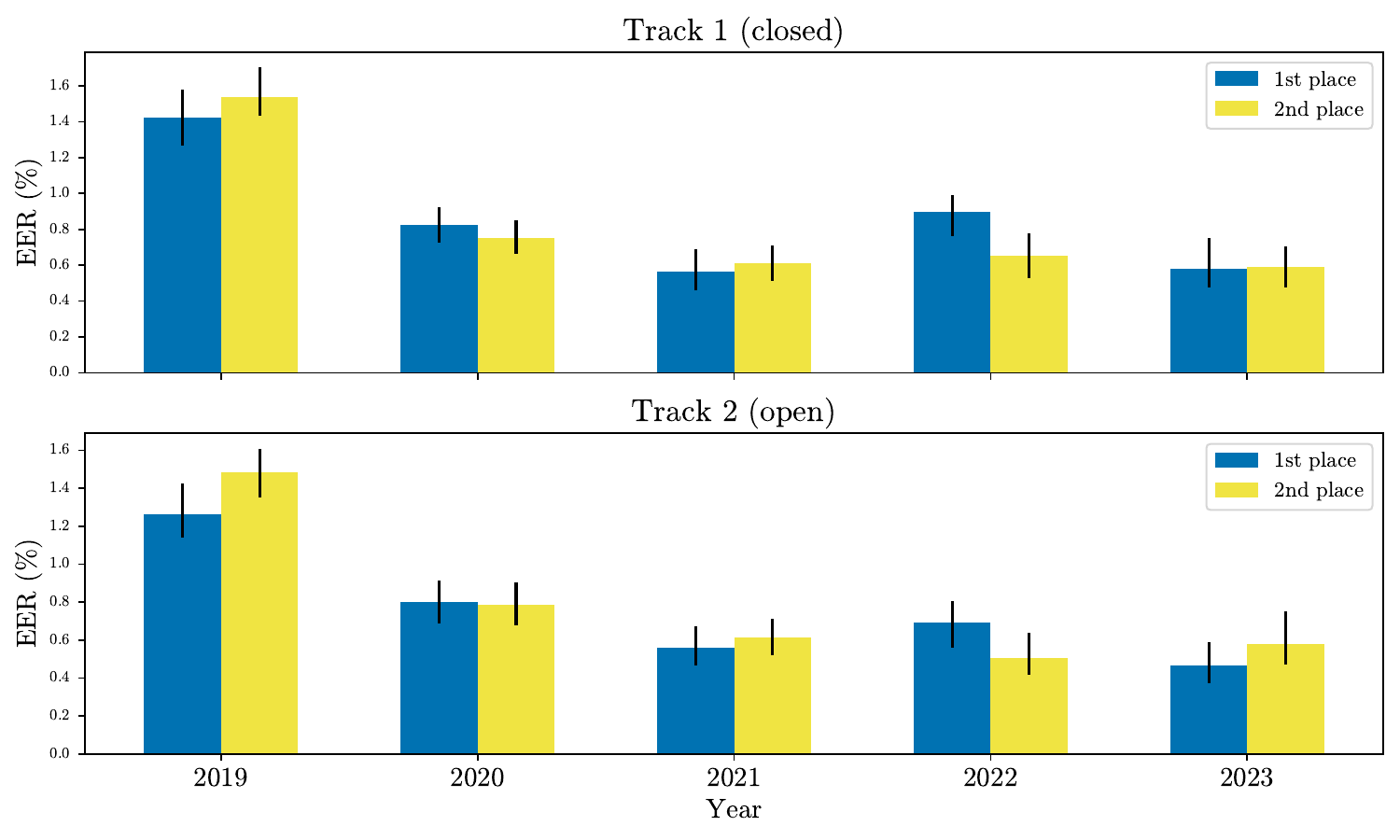}
    \caption{Performance confidence intervals (95\%) for the first and second places in each year of the VoxSRC2019 test set. The top figure shows the Track 1 submissions and the bottom figure shows the Track 2 submissions.}
    \label{fig:confidence_interval}
\end{figure}

We also computed the 95\% confidence intervals for the winning submissions on the VoxSRC 2019 test set. Figure \ref{fig:confidence_interval} shows the results.
To obtain these intervals, we adopted the conventional sample bootstrap approach, as described in Section 3.1.1 of \cite{poh2007estimating}: we generated 1000 samples with replacement for each submission and computed 1000 EER values.
Then, we calculated the quantiles corresponding to the 95\% confidence interval.
We decided to use the conventional bootstrap method instead of the user-specific bootstrap subset method for two reasons : (i) we want to preserve the label distribution during sampling and (ii) it aligns with our workshop setting -- we accept only one submission for the entire test set, not per enrolled speaker.
The minimum, average, and maximum widths of the confidence intervals are 18.6\%, 23.6\%, and 31.2\% of the EER value, respectively.
The confidence intervals demonstrate the variability in performance across different submissions, highlighting the importance of considering uncertainty in EER measurements. 
This is particularly relevant given the close performance between first and second place submissions in both tracks.

\subsection{Self-supervised speaker verification (2020-2021)}
\label{subsec:result_self_track3}
The self-supervised speaker verification track was held for two editions, VoxSRC 2020 and 2021.
The winners of this track employed a similar set of stages in both years: they (i) first trained the network using contrastive learning, (ii) then generated pseudo-labels based on the model from the first stage, and (iii) finally trained the network in a supervised way using these pseudo-labels. Table~\ref{tab:winner_comparison_self} displays the overall results.

 \begin{table}[t]
     \caption{Comparison of methods on test sets of self-supervised track. \linebreak We compare top-2 submissions of the year 2020 and 2021 on the 2019 test set. For \% EER shown, lower is better.}
     \vspace{-5pt}
    \label{tab:winner_comparison_self}
    \centering
    \resizebox{1\linewidth}{!}{
    \begin{tabular}{lcccc}
    \toprule
         \textbf{Team} & \textbf{Track} & \textbf{2019 test} & \textbf{2020 test}   & \textbf{2021 test}\\ \midrule
         Baseline~\cite{huh2020augmentation} & - & 12.67 & 29.93 & 19.74 \\ \midrule
          VoxSRC 2020 1st place~\cite{thienpondt2020idlab} & 3 & 2.26 &  \textbf{7.21} & -\\  
          VoxSRC 2020 2nd place~\cite{wang2020dku} & 3 & 6.49 & 12.42 & - \\
         \midrule
         VoxSRC 2021 1st place~\cite{cai2021dkudukeece} & 3 & \textbf{1.49} & - & \textbf{5.59} \\  
          VoxSRC 2021 2nd place~\cite{slavivcek2021phonexia} & 3 & 2.40 & - & 6.59 \\
          \bottomrule
    \end{tabular}}
    \vspace{-5pt}
\end{table} 

The VoxSRC 2020 winner~\cite{thienpondt2020idlab} exploited Momentum Contrast (MoCo)~\cite{he2020momentum} for the first stage, followed by iterative clustering using both efficient mini-batch k-means and Agglomerative Hierarchical Clustering (AHC) to make pseudo-speaker labels. 
A large ECAPA-TDNN was then trained with these labels using a sub-center AAM-softmax~\cite{Deng2020SubcenterAB} layer. 

The VoxSRC 2021 winner~\cite{cai2021dkudukeece} extended their previous two-stage iterative labelling framework~\cite{cai2021iterative}, which came second in 2020.
They also leveraged visual data on top of the audio data for the first time in the VoxSRC challenges.~\footnote{This is permitted for the self-supervised track.} 
They devised a unique clustering ensemble technique to fuse pseudo-labels from various modalities, which enhances the robustness of the speaker representations extracted.

Since we used the same test set across Tracks 1, 2 and 3, the self-supervised test set also includes the VoxSRC 2019 test set in both years 2020 and 2021.
Therefore, we can show the performance progress of the winners' method each year by measuring the performance on the VoxSRC 2019 test set.
Table~\ref{tab:winner_comparison_self} shows the result.
Compared to 2020, both the first and second place in VoxSRC 2021 show lower EER on VoxSRC 2019 test set, 2.26\% to 1.49\% and 6.49\% to 2.40\% respectively.
The winner of VoxSRC 2021 demonstrated a performance (EER 1.49\%) remarkably similar to the VoxSRC 2019 supervised track winner (EER 1.42\% on Track 1). 
This highlights the significant advancement of self-supervised methods in the field.

Note, we only hosted this track until 2021 for two reasons:
first, because most participants used similar techniques, adapting self-supervised methods from the computer vision literature; and second, because the scenario was
somewhat artificial (assuming that there were no labelled data) and we wished to move to the more practical scenario of semi-supervised domain adaptation, described next.

\subsection{Semi-supervised domain adaptation (2022-2023)}
\label{subsec:result_semi_track3}
The semi-supervised domain adaptation track is a new track introduced in VoxSRC 2022. 
Here, we describe the winners' submissions in the years 2022 and 2023.
Each year's winner performance is reported in Table~\ref{tab:winner_comparison_semi}.
Baseline~\cite{kwon2021ins} is the same baseline we used for Tracks 1 and 2, which is only trained with VoxCeleb2 dev set.

 \begin{table}[t]
 \vspace{-5pt}
     \caption{Comparison of methods on test sets of semi-supervised track. \linebreak We compare top-2 submissions of the year 2022 and 2023. For \% EER shown, lower is better.}
     \vspace{-5pt}
    \label{tab:winner_comparison_semi}
    \centering
    \resizebox{1\linewidth}{!}{
    \begin{tabular}{lccc}
    \toprule
         \textbf{Team} & \textbf{Track} & \textbf{2022 test}   & \textbf{2023 test}\\ \midrule
         Baseline~\cite{kwon2021ins} & - & 16.88 & 13.11 \\ \midrule
          VoxSRC 2022 1st place~\cite{zhao2023hccl} & 3 & \textbf{7.03} &  - \\  
          VoxSRC 2022 2nd place~\cite{qin2022dku} & 3 & 7.14 & - \\
         \midrule
         VoxSRC 2023 1st place~\cite{li2023dku} & 3 & - & \textbf{4.95} \\  
          VoxSRC 2023 2nd place~\cite{xx205_2023} & 3 & - & 8.13 \\
          \bottomrule
    \end{tabular}}
    \vspace{-5pt}
\end{table} 

The winner~\cite{zhao2023hccl} in VoxSRC 2022 used two frameworks, pseudo labelling and self-supervised learning.
A novel sub-graph clustering algorithm was used to generate pseudo-labels based on two Gaussian fitting and multi-model voting. 
The model was trained in two stages, first using the labelled source domain data and the pseudo-labelled target domain data, and secondly fine-tuning the CNCeleb data by fixing the VoxCeleb weights of the classification layer using circle loss. 
The pseudo-label correction method was then applied and the model was retrained with these new pseudo-labels. 
They also explored various types of domain adaptation techniques, such as CORAL~\cite{sun2017correlation} or CORAL+~\cite{lee2019coral+}.

The VoxSRC 2023 winner~\cite{li2023dku} introduced a novel pseudo-labelling method based on triple thresholds.
They utilised the well-trained speaker model using the source domain to extract embeddings from the target domain.
They conducted initial clustering using the K-Nearest Neighbours (KNN) algorithm, followed by data cleaning, sub-centre purification and class merging to obtain the pseudo labels.
At the last stage, they finetuned the model using both unlabelled data with pseudo-labels and labelled data.

\subsection{Speaker diarisation -- Track 4}
\label{subsec:result_track4}


\begin{table*}[t]
\caption{An overview of the winning methods on Track 4 across VoxSRC 2020 to 2023.}
\label{tab:diar_winner_method}
\resizebox{\textwidth}{!}{%
\begin{tabular}{@{}clllll@{}}
\toprule
\textbf{Workshop}                                              & \multicolumn{1}{c}{\textbf{VAD}}                                                                           & \multicolumn{1}{c}{Embedding network architecture} & \multicolumn{1}{c}{Clustering}                                        & \multicolumn{1}{c}{Fusion method} & \multicolumn{1}{c}{Other models used}                                         \\ \midrule
\textbf{\begin{tabular}[c]{@{}c@{}}VoxSRC\\ 2020\end{tabular}} & Conformer~\cite{gulati2020conformer} based Speech Separation                                                                          & Res2Net~\cite{gao2019res2net} variants                                   & - AHC                                                                 & Dover-LAP~\cite{raj2021dover}                         & - leakage filtering                                                           \\ \midrule
\textbf{\begin{tabular}[c]{@{}c@{}}VoxSRC\\ 2021\end{tabular}} & ResNet + BiLSTM                                                                                            & ResNet34 variant                                          & \begin{tabular}[c]{@{}l@{}}- AHC\\ - Spectral Clustering~\cite{lin2019lstm}\end{tabular} & Dover-LAP~\cite{raj2021dover}                           & \begin{tabular}[c]{@{}l@{}}- TS-VAD~\cite{Medennikov2020} \\ - Overlap speech detection\end{tabular} \\ \midrule
\textbf{\begin{tabular}[c]{@{}c@{}}VoxSRC\\ 2022\end{tabular}} & \begin{tabular}[c]{@{}l@{}}- ResNet based VAD\\ - Conformer~\cite{gulati2020conformer} based VAD\\ - pyannote.audio 2.0~\cite{bredin2020pyannote}\\ - ASR based VAD\end{tabular} & SimAM-ResNet34~\cite{qin2022simple}                                     & \begin{tabular}[c]{@{}l@{}}- AHC\\ - Spectral Clustering~\cite{lin2019lstm}\end{tabular} & Dover-LAP~\cite{raj2021dover}                           & - TS-VAD~\cite{Medennikov2020}                                                                    \\ \midrule
\textbf{\begin{tabular}[c]{@{}c@{}}VoxSRC\\ 2023\end{tabular}} & \begin{tabular}[c]{@{}l@{}}- ResNet based VAD\\ - Conformer~\cite{gulati2020conformer} based VAD\\ - ECAPA-TDNN~\cite{desplanques2020ecapa} based VAD \end{tabular} & SimAM-ResNet34~\cite{qin2022simple} + ResNet101 + SimAM-ResNet100 & - AHC & Dover-LAP~\cite{raj2021dover}                           & - Seq2Seq-TSVAD~\cite{cheng2023target}                                                                    \\ \bottomrule
\end{tabular}}
\end{table*}

\begin{table}[t]
    \centering
    \caption{Comparison of methods on test sets of the diarisation track. \linebreak We compare top-2 submissions of the year 2021, 2022, and 2023. \linebreak For \% DER shown, lower is better.}
    \label{tab:results_comparison_diarisation}
    \resizebox{1\linewidth}{!}{
    \begin{tabular}{lcccc}
    \toprule
     \textbf{Team} & \textbf{2021 test}&\textbf{2021 test} & \textbf{2022 test}& \textbf{2023 test} \\
     \midrule
     Baseline~\cite{simplediarization} & 8.15& 7.57 & 8.94 & 8.54 \\
    \midrule
        VoxSRC 2020 1st place~\cite{xiao2020microsoft}&      \textbf{3.71}           &    -      &    -   & -\\
         VoxSRC 2020 2nd place~\cite{landini2020analysis}& 4.00        &   -      &    -   &- \\
         \midrule
        VoxSRC 2021 1st place~\cite{wang2021dku}&     -      &         5.07      &           -   & -  \\
         VoxSRC 2021 2nd place~\cite{wang2021bytedance}&  -   &          5.15    &      -  &  -  \\
         \midrule
        VoxSRC 2022 1st place~\cite{wang2022dku}&      -      &    4.16    &     4.75            &  -  \\
         VoxSRC 2022 2nd place~\cite{cai2022kriston}& -     &   4.05 &    4.87    &  -   \\
        \midrule
        VoxSRC 2023 1st place~\cite{cheng2023dku}&      -      &    \textbf{3.74}    &     \textbf{3.94}      & \textbf{4.30}   \\
         VoxSRC 2023 2nd place~\cite{karamyan2023krisp}& -     &   4.06      &  4.23    & 4.71    \\
     \bottomrule
    \end{tabular}}
\end{table} 

The speaker diarisation track has been held since 2020.
Table~\ref{tab:diar_winner_method} compares the winning methods on Track~4.
The winning methods in this track all consist of the following steps: Voice Activity Detection (VAD) to detect the voice regions, speaker embedding extraction by using a sliding window approach over the voice region, and a clustering step to determine the speaker labels.

Winners used speech separation-based VAD or train their own VAD using ResNet~\cite{he2016deep}, Conformer~\cite{gulati2020conformer}, or ECAPA-TDNN~\cite{desplanques2020ecapa} architecture.
External public VAD models, such as the segmentation model from pyannote~\cite{bredin2020pyannote} or word-level timestamps from Kaldi ~\cite{povey2011kaldi} ASR systems were also explored.

Several embedding networks have been used such as variants of Res2Net~\cite{gao2019res2net} or SimAM-ResNet34~\cite{qin2022simple} trained with VoxCeleb. 
VoxSRC 2022 winner~\cite{wang2022dku} used pseudo-labels from VoxConverse utterances to mitigate the effect of domain mismatch between VoxCeleb and VoxConverse and VoxSRC 2023 winner~\cite{cheng2023dku} included additional data during training.

All winners used AHC to cluster the embeddings extracted using the sliding window approach.
Some winners~\cite{wang2021dku, wang2022dku} also adopted spectral clustering~\cite{lin2019lstm}.
After these three steps, all winning methods adopted Dover-LAP~\cite{raj2021dover} to fuse the results from different models.

Multiple methods were used to deal with overlapping speech.
Winners either trained a separate Overlap Speech Detection (OSD) model on their own or used TS-VAD~\cite{cheng2023target} to detect the overlapping speech regions and integrated the results into their final submission.

\noindent\textbf{Performance progression.} Table~\ref{tab:results_comparison_diarisation} shows the winners' performance on Track 4 test sets.
Since the first year's diarisation test set has not been included in the next three years, we cannot directly compare the VoxSRC 2020 winner with the winners in 2021, 2022 and 2023.
However, we included the VoxSRC 2021 test set in the VoxSRC 2022 and 2023 test set, and thus, we compare the performances between the last three years.
We could observe a steady performance improvement each year (DER of 5.07\% vs 4.05\% vs 3.74\%).

\section{Detailed analysis}
\label{sec:analysis}

In this section, we further analyse winners' submissions. 
We add analysis on (i) the verification pairs that winners get wrong, (ii) multi-lingual pairs in the year 2021, (iii) hard positive and negative pairs in the year 2022, and (iv) semi-supervised domain adaptation submissions with several baseline results in the year 2022.

\subsection{Verification pairs that winners get wrong}
\label{subsec:analysis_veripairs}
Over the previous five years, the state-of-the-art performance on the VoxSRC 2019 test set has progressively improved, as shown in Figure~\ref{fig:veri_progress}.
Models however still predict differently from the ground truth in some trials.

Of the 232 pairs that all the winners of VoxSRC 2022 and 2023 (first and second places on both closed / open tracks) get wrong,
16 are positive pairs and the rest are negative pairs.
We manually verify whether these pairs are correctly labelled or not.

For the positive pairs, 12 out of 16 are label errors.\footnote{We fixed these errors and released the new test labels in our official website~\url{https://mm.kaist.ac.kr/datasets/voxceleb/voxsrc/workshop.html}.}
Four of these errors are the result of utterances from different individuals with the same name, a confusion caused by the YouTube scraping process based on video titles. 
The remaining eight pairs are utterances from siblings who are mistakenly assigned the same identity due to their inclusion in YouTube videos titled with the name of a well-known celebrity.
These were not caught during the manual verification stage. 
Nevertheless, the proportion of erroneous labels is minimal and does not significantly impact the system's performance, accounting for only 0.006\% of the total 208,008 pairs.
The rest of the four pairs without label errors are from the same person but extracted at different ages, which is a challenging setting that we focused on in VoxSRC 2022, and is still an open area of research.

In the set of 216 negative pairs, no labelling errors are observed. 
This can be attributed to our method of random sampling from the speaker pools in the raw data, which significantly reduces the likelihood of labelling errors. 
All these pairs consist of different individuals of the same gender, and we do not notice any other significant trends.

\subsection{Analysis of performance by utterance length and gender}
\label{subsec:analysis_lengthgender}
\begin{table}[t]
\caption{Analysis of performance by utterance duration. \textbf{\textgreater x sec\ } denotes the subset of the VoxSRC2019 test pairs where both utterances are longer than x seconds. We report the performance of the VoxSRC2023 Track1 and Track2 winners.}
\label{tab:length_analysis}
\resizebox{\linewidth}{!}{%
\begin{tabular}{@{}lcccccccc@{}}
\toprule
\textbf{Eval set}                  & \multicolumn{2}{c}{\textbf{All}}     & \multicolumn{2}{c}{\textbf{\textgreater 4 sec}} & \multicolumn{2}{c}{\textbf{\textgreater 6 sec}} & \multicolumn{2}{c}{\textbf{\textgreater 8 sec}} \\ \midrule
\textbf{\# pairs}                  & \multicolumn{2}{c}{208,008} & \multicolumn{2}{c}{199,680}            & \multicolumn{2}{c}{51,621}             & \multicolumn{2}{c}{16,576}             \\ \midrule
                   & \textbf{EER}    & \textbf{minDCF}    & \textbf{EER}          & \textbf{minDCF}         & \textbf{EER}          & \textbf{minDCF}         & \textbf{EER}          & \textbf{minDCF}         \\ \midrule
\textbf{Track 1 winner~\cite{zheng2023unisound}} & 0.58            & 0.028              & 0.58                  & 0.027                   & 0.37                  & 0.019                   & 0.11                  & 0.008                   \\
\textbf{Track 2 winner~\cite{torgashov2023id}} & 0.47            & 0.020              & 0.45                  & 0.019                   & 0.26                  & 0.014                   & 0.03                  & 0.006                   \\ \bottomrule
\end{tabular}}
\end{table}

In this section we analyse the performance on the VoxSRC2019 test set of the Track 1~\cite{zheng2023unisound} and Track 2~\cite{torgashov2023id}  VoxSRC2023 winners against utterance length and gender.

Table~\ref{tab:length_analysis} shows the performance as the utterance length is varied, where
\textbf{\ \textgreater x sec\ } denotes the subset of VoxSRC2019 test pairs where both utterances of each pair are longer than x seconds. 
Both models show better performance as the utterance length increases. This is 
because there is a greater chance that relevant speech signals from the actual speaker are captured, as has been studied in a number of works~\cite{vogt2009minimising,Xie19a}.
Interestingly, the Track 2 winner shows an impressive 0.03\% on EER on the \textbf{\ \textgreater 8 sec\ } set --
only five out of 16,576 pairs are predicted incorrectly at the EER point when the false positive rate and false alarm rate are the same.

We provide an analysis of the performance as a function of gender in Table~\ref{tab:gender_analysis}. \textbf{Male} and \textbf{Female} denote the subsets of the VoxSRC2019 test pairs where both utterances are from male and female, respectively. 
For both models, performance is better on the male set than the female set.
We assume that this is due to the gender imbalance of the VoxCeleb2 dev set, which is used as our training set : it consists of 61\% of male and 39\% of female~\cite{Chung18a}. However, since both models already show strong performance, the performance gap between gender is not huge. (0.12\% and 0.16\% for 1st and 2nd place, respectively)

\begin{table}[t]
\caption{Analysis of performance by gender. \textbf{Male} and \textbf{Female} denote the subsets of the VoxSRC2019 test pairs where both utterances are from male and female, respectively. We report the performance of the VoxSRC2023 Track1 and Track2 winners.}
\label{tab:gender_analysis}
\resizebox{\linewidth}{!}{%
\begin{tabular}{@{}lcccccc@{}}
\toprule
\textbf{Eval set}       & \multicolumn{2}{c}{\textbf{All}}     & \multicolumn{2}{c}{\textbf{Male}}   & \multicolumn{2}{c}{\textbf{Female}} \\ \midrule
\textbf{\# pairs}       & \multicolumn{2}{c}{208,008} & \multicolumn{2}{c}{85,738} & \multicolumn{2}{c}{78,750} \\ \midrule
\textbf{}               & \textbf{EER}    & \textbf{minDCF}    & \textbf{EER}    & \textbf{minDCF}   & \textbf{EER}    & \textbf{minDCF}   \\ \midrule
\textbf{Track 1 winner~\cite{zheng2023unisound}} & 0.58            & 0.028              & 0.57            & 0.026             & 0.69            & 0.034             \\
\textbf{Track 2 winner~\cite{torgashov2023id}} & 0.47            & 0.020              & 0.43            & 0.018             & 0.59            & 0.025             \\ \bottomrule
\end{tabular}}
\end{table}

\subsection{Multi-lingual focus -- VoxSRC 2021}
\label{subsec:analysis_multilingual}
The verification tracks in 2021 had a multi-lingual focus, via the inclusion of multi-lingual data in the test set. 
We provide some analysis on the performance of the winning methods from the supervised tracks (1st and 2nd places on Track 1) and the provided baseline~\cite{chung2020defence} on the multi-lingual data.

\begin{figure}[t]
\centering
        \includegraphics[width=0.95\linewidth]{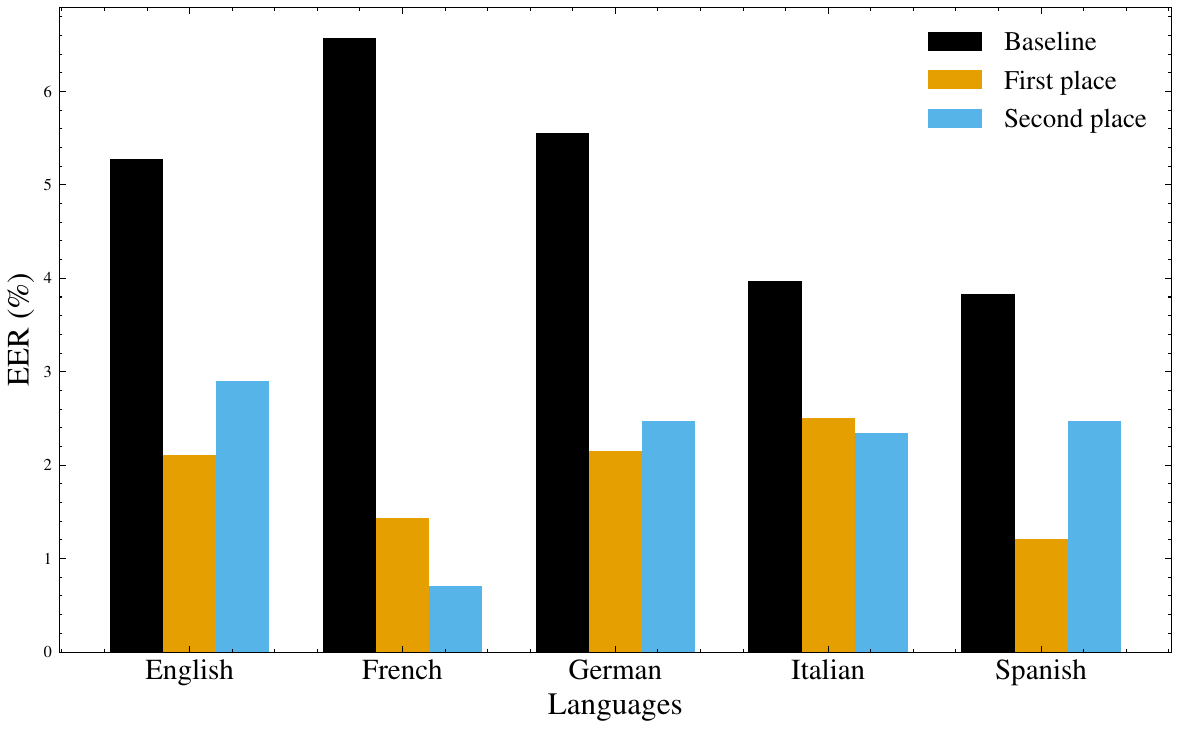}
          \caption{The performances of different models on the pairs only with certain languages in the VoxSRC 2021 test set.}
        \label{fig:withinlang}
\end{figure}

We measure performance on the identical language subsets for the five most common languages in VoxCeleb. 
The results are shown in Figure~\ref{fig:withinlang}. 
We calculate a performance measure per language by calculating the EER from the same-language pairs.  
For all five languages, there are at least 1000 same-language pairs (both positive and negative) in the test set. 
The exact number of same-language pairs per language is: English: 282,039; French: 5,987; German: 1,066; Italian: 1,129; and Spanish: 4,241.
The winning methods perform much better than the baseline across all languages, although there is still considerable variation in performance between different languages.
Interestingly, although English is the most common language in the training set, none of the models performs best on the English language pairs. 
However, we note that there is more statistical uncertainty for the languages with a lower number of pairs, \textit{e.g.\ }German with 1,066 pairs, compared to those with many pairs, \textit{e.g.\ }English with 282,039 pairs.
This is due to the error introduced when estimating population statistics using small sample sizes.

\subsection{Hard positive and negative pairs -- VoxSRC 2022}
\label{subsec:analysis_hard}

In the year 2022, we introduced new trial types to make the test set harder -- hard positive pairs with large age gaps, and hard negative pairs that share the same background noise.
Here we analyse how these pairs affect the winners' performance.
The VoxSRC 2022 test set consists of four types of trials, (i) hard positive pairs taken from the same speaker at different ages (\textbf{P-Hard}), (ii) hard negative pairs taken from the same environment (\textbf{N-Hard}), (iii) positive pairs from VoxSRC 2019 test set (\textbf{P-Vox19}), and (iv) negative pairs from VoxSRC 2019 test set (\textbf{N-Vox19}).
We compare the performance of our baseline model~\cite{chung2020defence} and the top 2 winners of Track 1 on these subsets.

Table~\ref{tab:analysis_2022} shows the results. 
The 1st~\cite{ravana2022} and 2nd place~\cite{cai2022kriston} performed better than our baseline model by a large margin. 
Comparing the performance of E-1 to the others shows that both the hard positives and the hard negatives made the challenge more difficult.
For the most challenging set, E-4 with both hard positive and negative pairs, the 1st place method (which achieves an impressive 0.9 \% EER on the VoxSRC-19 test set) could only achieve 2.07\%.
Interestingly, the second-place method performed better in E-1, E-2 and E-3 than the 1st place but achieved worse results in E-4.
This led to a slightly better EER performance of the second place in 2022 than the first place on the overall VoxSRC2022 test set (1.40\% vs 1.49\%). 
While both use similar training strategies and identical training set in Track 1, the first place uses ResNet100 with a base channel of 128, while the second place uses the variants of ResNet34 with a base channel of 64. The larger capacity of the first place might have led to a slight overfitting to the training set. However, as shown in Table~\ref{tab:analysis_2022}, the difference is not significant.

\begin{table}[!t]
    \centering
        \caption{Performance of baseline model and winning methods in VoxSRC 2022 Track 1 on four subsets of the test set. Reported in EER (\%). }
    \resizebox{\linewidth}{!}{%
    \begin{tabular}{cccccc}
    \toprule
    \textbf{Eval set} & \textbf{Positive pairs} & \textbf{Negative pairs} & \textbf{Baseline~\cite{chung2020defence}} & \textbf{1st place~\cite{ravana2022}} & \textbf{2nd place~\cite{cai2022kriston}} \\ \midrule
        \textbf{E-1} & P-Vox19 & N-Vox19 & 1.47  &  0.90  & \textbf{0.65} \\
        \textbf{E-2} & P-Vox19 & N-Hard & 3.25  &  1.35  & \textbf{1.15} \\
        \textbf{E-3} & P-Hard & N-Vox19 & 4.50  &  1.33  & \textbf{1.18} \\
        \textbf{E-4} & P-Hard & N-Hard & 9.27  &  \textbf{2.07}  & 2.28 \\
    \bottomrule
    \end{tabular}}
    \label{tab:analysis_2022}
\end{table} 

\subsection{Semi-supervised domain adaptation -- VoxSRC 2022}

Table~\ref{tab:track3_analysis} shows the top two teams' performance on the test set compared to the baselines. 
\textbf{Baseline~1} is the model which is trained only with the VoxCeleb2 dev set, the labelled data in the \textit{source} domain (\textbf{L-S}).
We also provide \textbf{Baseline~2}, which is trained only with the labelled data in the \textit{target} domain (\textbf{L-T}) from scratch. 
\textbf{Baseline~3} is trained to start from Baseline 1 and finetuned with labelled data in the target domain using a low learning rate (1e-5).
All baseline models are ResNetSE34V2~\cite{kwon2021ins} with ASP pooling~\cite{okabe2018attentive} and are trained with a combination of angular prototypical loss~\cite{chung2020defence} and cross-entropy loss.
Neither of these baselines takes advantage of the large amount of unlabelled target domain data available to participants.

A comparison of Baselines 1 and 3 shows that including the labelled data in the target domain results in a performance improvement, 2.95\% in terms of EER, even though the size of the labelled target domain data is negligible. 
However, Baseline~2 shows that using only the labelled target domain data results in a substantial performance decrease due to overfitting. 
Finally, the two winners' performances show that utilising the extensive unlabelled target domain data is essential for performance improvement in the training set, such as in the form of pseudo-labelling during training. 

 \begin{table}[!ht]
     \caption{Comparison of winning methods in Track3 with baselines. \linebreak \textbf{L-S} : Labelled data in Source domain, \textbf{U-T}: Unlabelled data in Target domain and \textbf{L-T} Labelled data in Target domain.}
    \label{tab:track3_analysis}
    \centering
    \begin{tabular}{cccc}
    \toprule
         \textbf{Model} & \textbf{Train dataset} & \textbf{minDCF} &\textbf{EER}  \\ \midrule
        Baseline 1 & \textbf{L-S} & 0.823 & 16.88 \\
        Baseline 2 & \textbf{L-T} & 0.999 & 32.47 \\
        Baseline 3 & \textbf{L-S} + \textbf{L-T} & 0.687 & 13.93\\
        \midrule
        1st place~\cite{zhao2023hccl} & \textbf{L-S} + \textbf{U-T} + \textbf{L-T} & \textbf{0.388} & \textbf{7.03} \\
        2nd place~\cite{qin2022dku} & \textbf{L-S} + \textbf{U-T} + \textbf{L-T} & 0.389 & 7.15 \\
    \bottomrule
    \end{tabular}
\end{table} 
\section{Workshop}
\label{sec:workshop}

\begin{table*}[t]
\caption{Workshop information}
\label{tab:workshop_stats}
\resizebox{1\linewidth}{!}{
\begin{tabular}{@{}clll@{}}
\toprule
\textbf{}                                                       & \multicolumn{1}{c}{\textbf{Date}} & \multicolumn{1}{c}{\textbf{Location}} & \multicolumn{1}{c}{\textbf{Keynote speech}}                                                                                                                                                                                                                  \\ \midrule
\textbf{\begin{tabular}[c]{@{}c@{}}VoxSRC\\ 2019\end{tabular}}  & Sep 14th, 2019                    & Graz, Austria                      & Mitchell McLaren, "Speaker recognition - a retrospective"                                                                                                                                                                                                       \\
\textbf{\begin{tabular}[c]{@{}c@{}}VoxSRC \\ 2020\end{tabular}} & Oct 30th, 2020                    & Virtual                            & \begin{tabular}[c]{@{}l@{}}Daniel Garcia-Romero, "X-vectors: Neural Speech Embeddings for Speaker Recognition"\\ Shinji Watanabe, "Tackling Multispeaker Conversation Processing based on Speaker Diarization and Multispeaker Speech Recognition"\end{tabular} \\
\textbf{\begin{tabular}[c]{@{}c@{}}VoxSRC \\ 2021\end{tabular}} & Sep 7th, 2021                     & Virtual                            & Andreas Stolcke, "Speaker Recognition and Diarization for Alexa"                                                                                                                                                                                                \\
\textbf{\begin{tabular}[c]{@{}c@{}}VoxSRC \\ 2022\end{tabular}} & Sep 22nd, 2022                    & Incheon, Korea                     & Junichi Yamagishi, "The use of speaker embeddings in neural audio generation"                                                                                                                                                                                   \\ 
\textbf{\begin{tabular}[c]{@{}c@{}}VoxSRC \\ 2023\end{tabular}} & Aug 20th, 2023                    & Dublin, Ireland                    & Wei-Ning Hsu, “Scalable controllable speech generation via explicitly and implicitly disentangled speech representations”                                                                                                                                        \\ \bottomrule
\end{tabular}}
\end{table*}

VoxSRC workshops were held either in-person or virtually as a Zoom webinar in conjunction with the Interspeech conference.
The workshops were open for all to attend without any charge. 
They typically started with an overview of the competition, followed by a keynote speech, an announcement of winners, and presentations by the winners. 
The winners were selected only from teams that submitted technical reports to the organisers.
Details such as dates, locations, and the keynote speaker each year can be found in Table~\ref{tab:workshop_stats}.
For the winners’ talks, the virtual attendees sent the organisers pre-recorded videos which explained their methods and results, while the in-person attendees gave their talks in the workshop venue, if possible.
Questions were gathered from both the virtual participants as well as those present physically at the venue.

\begin{figure}[t]
 \centering
        \includegraphics[width=\linewidth]{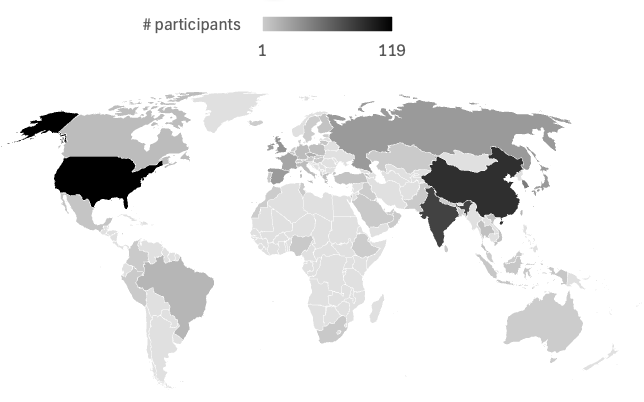}
          \caption{Geographical heatmap of virtual participants from 2020 to 2023. \textbf{\# participants} denotes the number of participants who attended the workshops virtually per country.}
        \label{fig:geoheatmap}
\end{figure}

Due to the COVID-19 pandemic, since 2020 we have sent a Zoom link to those who registered for our workshop through EventBrite~\cite{eventbrite}.
According to statistics gathered from EventBrite, 
over the last four years, 776 people in total registered for the workshop virtually from 59 different countries.
Figure~\ref{fig:geoheatmap} shows the geographical heatmap of virtual participants per country from 2020 to 2023. The top 5 countries where participants have come from are the United States (119 participants), followed by China (92), India (81), Korea (41) and the United Kingdom (34). 

\section{Discussion and the Future}
\label{sec:discussion}
In this section, we present an analysis of the trends and constraints of the current generation of algorithms, as observed in our workshops over the past five years. 
Additionally, we discuss the specific limitations encountered in our workshop setting. 
The section concludes with key insights and recommendations intended for future organisers who are interested in hosting similar challenges.

\subsection{Analysis of winners' methods}
\label{subsubsec:trends_winner}
Throughout the five series of VoxSRC challenges, there have been significant advances in speaker recognition technology, with 
the most recent speaker verification models easily achieving super-human performance (since humans are poor at recognising unfamiliar voices)~\cite{familiar_vs_unfamiliar_ch7, Greenberg2011IncludingHE}.
Models can now generate representative speaker embeddings that are robust to diverse conditions such as domain and language.
Speaker verification performance on VoxCeleb is now approaching 0\% EER, even though conditions are ``in the wild'' and much shorter (mostly under 10 seconds) than those of a decade ago (minutes or even hours). 
Speaker diarisation models have benefited from the progress in speaker verification, as a speaker embedding extractor is a core component of diarisation.  

What are the trends of the current generation of methods observed in our workshop? 
We consider the trends first for speaker verification and then for diarisation.

\noindent\textbf{Trends in verification models.}
Although the performance of speaker verification has improved significantly over the five years of the challenge, the underlying methodology remains largely unchanged compared to the approach of the VoxSRC 2019 winner. 
As detailed in Table~\ref{tab:veri_winner_method}, all winners follow a similar process: (i) inputting the voice as a spectrogram into a CNN-based model, (ii) training the CNN with AM-softmax or AAM-softmax loss, (iii) applying extensive data augmentation or using external data during training, and (iv) employing back-end systems like score normalisation or quality metric-based calibration methods to fuse results from multiple models. 
The components of this recipe, present in 2019 systems, have been refined and improved over the past five years -- for example, different backbones like TDNNs~\cite{waibel2013phoneme, peddinti2015time,desplanques2020ecapa} and Conformers~\cite{gulati2020conformer} can be used.

Some winners in Track 2~\cite{ravana2022, torgashov2023id} utilised self-supervised pretrained networks primarily for feature extraction, training an additional network with these features to be used in an ensemble.
Direct finetuning of self-supervised pretrained networks, however, has not led to performance gains.

This raises the question: Can self-supervised front-ends or transformers yield significant improvements as seen in other fields? 
Transformers have undoubtedly made remarkable progress in natural language processing~\cite{devlin2019bert, touvron2023llama} and computer vision~\cite{dosovitskiy2020vit, carion2020end, radford2021learning}. 
Yet, all winners adapted large ResNet~\cite{he2016deep}-based models for their submissions or used features from transformer-based models to train additional convolution-based networks.
It is acknowledged that transformers, lacking some inductive biases inherent in CNNs such as translation equivariance and locality, do not generalise well with limited data. 
However, with sufficient data, large-scale training can surpass these inductive biases. 
With the growing availability of public speaker verification data, it becomes feasible to train transformers leveraging these extensive datasets.
We leave this as a future direction in this field.

\noindent\textbf{Trends in diarisation models.}
In speaker diarisation, three distinct approaches are used: (i) the use of existing voice activity detection (VAD) combined with a speaker model and clustering~\cite{dgr2017dnn_diarization, wang2018speaker, zhang2019fully, kwon2021look}, (ii) the application of End-to-End Neural Diarisation (EEND), which progresses from VAD to speaker assignment~\cite{fujita2019end, horiguchi2020end}, and (iii) a combination of the strengths of these two methods by operating locally with end-to-end methods and then globally combining the intermediate results via clustering~\cite{bredin2023pyannote, cornell2023one}.

Interestingly, all winners in our workshop consistently chose the clustering-based approach (i). 
However, this approach inherently faces several limitations, such as difficulty in handling overlapping speech and the challenge of optimising various components for diarisation performance. 
In contrast, the EEND method effectively addresses overlapping speech by introducing multiple speaker labels for each timestamp.
It is also capable of directly optimising a single model for diarisation performance.

Of course, EEND also has its limitations: (i) it suffers when processing audio with a large number of speakers or long duration, (ii) it presents challenges for real-time processing as it struggles to match the speakers between the output of audio chunks during block-wise processing, and (iii) it is prone to overfitting to the training data distribution, as noted by~\cite{park2022review}. 
However, this direction could open up unprecedented opportunities to address challenges in the field of speaker diarisation, such as joint optimisation with speech recognition or speech separation, with overlapping speech.

In recent years, methods that combine EEND models with clustering techniques have gained interest~\cite{kinoshita2021integrating, bredin2023pyannote}. 
These methods apply the EEND model on a local scale to effectively manage overlapping speech while employing clustering methods that utilise global speaker embeddings derived from the EEND model.
The objective is to resolve the inter-block label permutation issue that arises during the block-wise processing of long audio recordings. 
This approach was employed in a recent submission~\cite{pyannoteVoxSRC2023} and we expect its further improvement in the future.

\subsection{Current research challenges}
Although the performance of speaker recognition models saturates on VoxSRC test sets, there still remain several open challenges for future speaker verification workshops.
This section describes outstanding problems that could form possible future challenges for speaker verification workshops.

\noindent\textbf{Anti-spoofing.} 
An important use case of speaker recognition is biometric verification. 
However, in real-world scenarios, e.g., online banking, there is a need to safeguard against spoofing attacks  -- including synthesized utterances, an area where performance is rapidly improving. 
Currently, it is well known that even state-of-the-art models are highly vulnerable when fed with spoofed inputs. 
The ASVSpoof~\footnote{https://www.asvspoof.org/} evaluation series is facilitating advancements in this important topic.

\noindent\textbf{Noisy or overlapping scenarios.}
Although VoxCeleb is ``in the wild'' to some extent, it still does not cover extremely noisy or overlapping scenarios.
With these hurdles remaining to be dealt with, we urge the research community to collect and curate a dataset beyond VoxCeleb, and even closer to real-world scenarios which include highly noisy and overlapping inputs.

\noindent\textbf{Diversity and scale in data.}
While VoxCeleb is recognised as one of the largest speaker verification datasets in the research community, and although it is multi-lingual, it is dominated by English speakers (approximately 84\% in VoxCeleb1 and 53\% in VoxCeleb2, based on the output of VoxLingua~\cite{valk2021slt} classifier).
In our semi-supervised domain adaptation track, we made efforts to include Chinese data, yet our dataset still lacks representation of more diverse languages, particularly low-resource languages. 
Additionally, the gender distribution in VoxCeleb2 is somewhat imbalanced, as discussed in Section~\ref{subsec:analysis_lengthgender}. 
Diversity in AI systems is important to ensure systems represent the variety of people who will use them.
Additionally, addressing ethical concerns and ensuring fairness in speaker recognition systems is critical. 
These issues require attention in any future dataset collection and release.

In terms of scale, while VoxCeleb contains approximately 2,000 hours of speech, the state-of-the-art ASR systems, such as Whisper~\cite{radford2023robust}, are typically trained on millions of hours of speech data.
It is well known that training with large amounts of data leads to better performing models.
It would be helpful for the community if challenge organisers could curate and provide large scale training data for participants to push the limits of their models.

\subsection{Lessons for future challenge organisers}
\label{subsec:lesson_organisers}

Organising a successful challenge in the field of data science or machine learning requires meticulous planning and foresight. 
Based on our experience, we note several key lessons that future challenge organisers should consider:

\noindent\textbf{Importance of a robust evaluation platform.} 
A critical factor in the success of any challenge is the reliability of the evaluation platform. 
The platform must be stable to handle a large volume of submissions and a diverse of data types.
Additionally, it should be easily adjustable to accommodate evolving challenge requirements and criteria. 
This flexibility ensures that the platform remains relevant and efficient throughout the challenge and beyond.
In recent times, the advent of numerous platforms~\cite{kaggle, EvalAI, codalab_competitions} for machine learning challenges has significantly aided organisers in hosting such events.
 We recommend that future organisers carefully select their platforms, taking into account the diverse features and functionalities that these platforms offer, and promised longevity.

\noindent\textbf{Maintain a persistent test set.}
Effectively tracking state-of-the-art performance annually is essential in hosting a series of machine learning challenges. 
This can be achieved by maintaining a consistent test set. 
There are two possible approaches: either keeping the entire test set unchanged or incorporating the test set from previous challenges into the subsequent year's test set.
This continuity makes it possible to directly compare the methods used by the winners each year and to analyse the progress made by state-of-the-art models.

\noindent\textbf{Secure non-overlapping test sets.}
 As the Internet continues to grow with ever-increasing volumes of data, the challenge is to curate test sets that are truly challenging and reflect real-world scenarios. 
 A particular challenge is siloing the test set, to ensure that it has not been used for training.
 Since 2020, the winners achieved below 1\% EER on the original VoxSRC 2019 test set,  which we consider as performance saturation. Consequently, each year of the challenge we developed more challenging test sets -- and this occupied most of the preparation time before a new challenge was launched.
Future organisers need to be innovative in the way they create harder datasets, ensuring that they test the boundaries of current methodologies. 
 
\noindent\textbf{Report individual and ensemble network performance.} 
It is proven that using an ensemble of several individual models leads to better generalisation performance~\cite{ganaie2022ensemble}. 
We observe that all challenge winners fuse predictions from multiple models in order to achieve better performance on the test set.
However, challenge reports should include detailed analyses of the performance of {\em individual} networks as well as their ensemble counterparts. 
Addressing both approaches provides a comprehensive view of the strengths and limitations of different models and offers insights into their practical applications. 

\noindent\textbf{More ...} We suggest that challenge organisers also read the advice from other challenges, such as the `Discussion and Future' section of the PASCAL VOC retrospective~\cite{Everingham15}.
\section{Conclusion}
\label{sec:conclusion}
This paper presents a review of the VoxSRC challenges and workshops including how we host the challenge, how we create the datasets and methods from challenge participants. 
We conclude with a discussion and limitations of our workshop along with our advice for future challenge organisers.
We hope this helps researchers in the speaker recognition and diarisation field to grasp an overall trend in these fields and also people who want to host a similar challenge in the future.

\section*{Acknowledgments}
First, we would like to thank our advisors,  Mitchell McLaren and Douglas A. Reynolds for providing valuable comments and feedback.
Second, we thank the annotators who helped us to make all datasets used. We especially thank 
Rajan from Elancer and his team for the annotation of the VoxConverse and diarisation test sets. We also thank Bong-Jin Lee, Heesoo Heo, Youngki Kwon, Youjin Kim, Triantafyllos Afouras, Alba Maria Martinez Garcia, Kihyun Nam, Doyeop Kwak and Youngjoon Jang for double-checking the diarisation validation and test set labels.
Third, we are grateful to those who helped maintain our evaluation servers including David Pinto, Ernesto Coto and Abhishek Dutta.
Fourth, we thank the CNCeleb authors for generously providing their hidden dataset as well as additional support.
We also thank Naver Corporation for sponsoring our workshops. 

{\appendices
\section{Statistics of validation and test sets}
\label{appendix:stats_valtest}
Table~\ref{tab:veri_stats} shows the statistics of the verification datasets and Table~\ref{tab:diar_stats} shows the statistics of the diarisation datasets
for the entire validation and test datasets over the five years of the challenge.
The verification datasets of the verification tracks up to 2021 are constructed using the entire VoxCeleb1, so \# of speakers is limited to 1,251, which is the number of speakers in VoxCeleb1.
The \# of speakers increased significantly in 2022 when we started to include the utterances of VoxConverse for hard negative pairs.

For the diarisation datasets, we started to use the entire VoxConverse set as our validation set, so the validation set statistics are identical from 2021 to 2023.
In 2020, there was a significant difference in the dataset distribution between the validation and test sets in terms of the average number of speakers and the average length of the audio.
We fixed this issue from 2021 on.

\begin{table}[!h]
\caption{Statistics of validation and test sets in the verification tracks. Note that validation sets of track 3 in year 2022 and 2023 are identical. \textbf{\# files} : total number of audio files, \textbf{\# pairs} : total number of pairs, \textbf{\# spks} : total number of speakers and \textbf{Avg length} : average length of audio in seconds.} 
\label{tab:veri_stats}
\centering
\resizebox{\linewidth}{!}{
\begin{tabular}{@{}lcrrrcrrrc@{}}
\toprule
\multirow{2}{*}{\textbf{Year}} & \multirow{2}{*}{\textbf{Track}} & \multicolumn{4}{c}{\textbf{Validation set}}                                                                                                & \multicolumn{4}{c}{\textbf{Test set}}                                                                                                      \\ \cmidrule(l){3-10} 
                               &                                 & \multicolumn{1}{c}{\textbf{\# files}} & \multicolumn{1}{c}{\textbf{\# pairs}} & \textbf{\# spks} & \multicolumn{1}{c}{\textbf{Avg length}} & \multicolumn{1}{c}{\textbf{\# files}} & \multicolumn{1}{c}{\textbf{\# pairs}} & \textbf{\# spks} & \multicolumn{1}{c}{\textbf{Avg length}} \\ \midrule
\textbf{2019}                  & 1,2                             & 4,715                                 & 37,720                                & 40               & 8.3                                     & 19,154                                & 208,008                               & 745              & 7.4                                     \\ \midrule
\textbf{2020}                  & 1,2                             & 140,815                               & 263,486                               & 1,251            & 8.2                                     & 118,439                               & 1,695,248                             & 1,440            & 5.0                                     \\ \midrule
\textbf{2021}                  & 1,2,3                           & 64,711                                & 60,000                                & 1,251            & 8.1                                     & 116,984                               & 476,224                               & 1,441            & 5.0                                     \\ \midrule
\multirow{2}{*}{\textbf{2022}} & 1,2                             & 110,366                     & 305,196                               & 2,307            & 8.4                                     & 34,684                                & 317,973                               & 1,974            & 7.4                                     \\ \cmidrule(l){2-10} 
                               & 3                               & 2,400                                 & 40,000                                & 120              & 9.4                                     & 18,377                                & 30,000                                & 56               & 9.1                                     \\ \midrule
\multirow{2}{*}{\textbf{2023}} & 1,2                             & 63,516                                & 49,987                                & 2,140            & 8.2                                     & 256,547                               & 825.437                               & 2,532            & 5.3                                     \\ \cmidrule(l){2-10} 
                               & 3                               & 2,400                                 & 40,000                                & 120              & 9.4                                     & 21,997                                & 80,000                                & 56               & 8.9                                     \\ \bottomrule
\end{tabular}}
\end{table}
\begin{table}[!ht]
\caption{Statistics of track 4 validation and test sets. \textbf{\# files} : total number of audio files, \textbf{Avg \# spks} : average number of speakers per each audio and \textbf{Avg length} : average length of audio in seconds.} 
\centering
\label{tab:diar_stats}
\resizebox{\linewidth}{!}{
\begin{tabular}{@{}lcccccc@{}}
\toprule
\multirow{2}{*}{\textbf{Year}} & \multicolumn{3}{c}{\textbf{Validation set}}                                                               & \multicolumn{3}{c}{\textbf{Test set}}                                                                    \\ \cmidrule(l){2-7} 
                      & \multicolumn{1}{c}{\textbf{\# files}}  & \multicolumn{1}{c}{\textbf{Avg \# spks}} & \multicolumn{1}{c}{\textbf{Avg length}} & \multicolumn{1}{c}{\textbf{\# files}} & \multicolumn{1}{c}{\textbf{Avg \# spks}} & \multicolumn{1}{c}{\textbf{Avg length}} \\ \midrule
2020                  & 216                           & 4.5                             & 338.2                          & 232                          & 6.5                             & 675.6                          \\ \midrule
2021                  & \multirow{3}{*}{448} & \multirow{3}{*}{5.5}            & \multirow{3}{*}{512.9}         & 264                          & 5.6                             & 452.1                          \\ \cmidrule(r){1-1} \cmidrule(l){5-7} 
2022                  &                               &                                 &                                & 360                          & 5.5                             & 449.2                          \\ \cmidrule(r){1-1} \cmidrule(l){5-7} 
2023                  &                               &                                 &                                & 413                          & 5.8                             & 519.6                          \\ \bottomrule 
\end{tabular}}
\end{table}
\section{Participant statistics}
\label{appendix:participant_statistics}
Figure~\ref{fig:participant_stats} shows a bar chart comparing the participation statistics per year over the course of the challenge.
The left figure shows the progress of the number of teams that participated each year, while the right shows the change of number of submissions each year.
The number of participants increased until 2021 and then decreased.
However, the number of submissions increased until 2022, which means that the number of people participating and submitting their results multiple times before the deadline has increased.
Interestingly, the number of participants in Track 2 increased dramatically from 2020 to 2021 (from 21 to 54), due to the emergence of the self-supervised speaker models~\cite{hsu2021hubert, chen2022wavlm}.

\begin{figure}[t]
\centering
    \includegraphics[width=\linewidth]{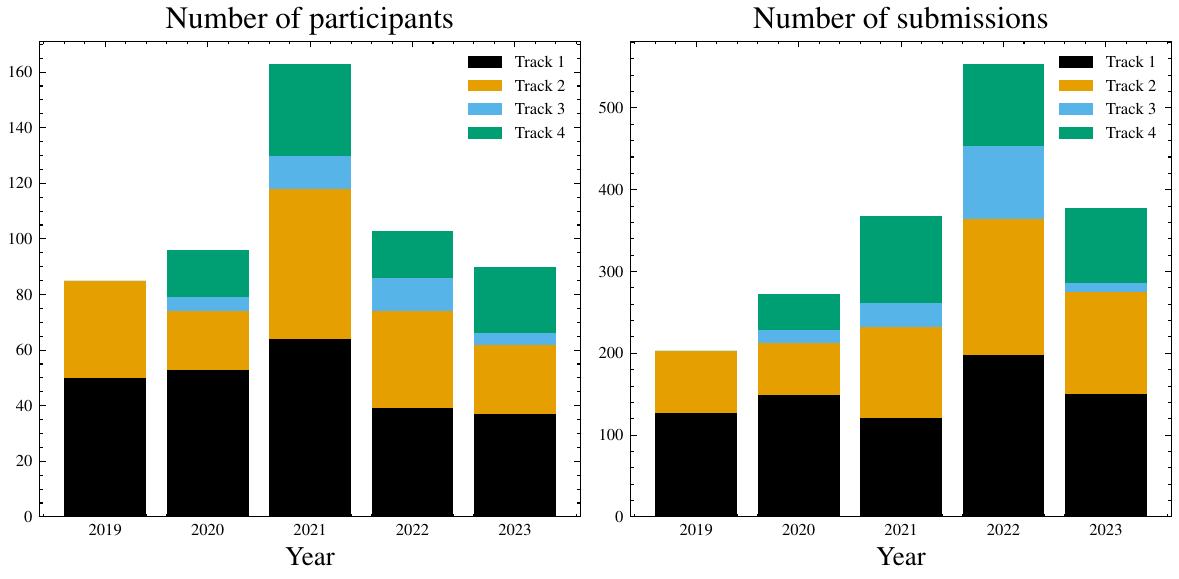}
      \caption{Number of participants (left) and number of submissions (right) over the series of the VoxSRC challenge.}
    \label{fig:participant_stats}
\end{figure}

\section{Potential risks and societal impacts}
One potential risk of this work relates to privacy concerns. 
Although identities have been anonymised, misuse or unauthorised access to this data could lead to privacy violations or identity theft. 
We advise researchers and developers using this dataset to be aware of these risks when training models with this data.
Furthermore, it is important to note that the distribution of identities in our dataset does not represent the global human population. 
We urge careful consideration of potential unintended biases -- including societal, gender, and racial biases -- when training or deploying models using this data for either training or evaluation purposes.
}

\bibliographystyle{IEEEtran}
\bibliography{shortstrings, mybib, vgg_local, vgg_other}

\section{Biography Section}

\vspace{-35pt}
\begin{IEEEbiographynophoto}{Jaesung Huh} is currently working towards DPhil degree in Visual Geometry Group, University of Oxford, supervised by Andrew Zisserman. Before that, he used to be a research engineer at Naver Corporation. His research focuses on audio-visual learning and video understanding. He was one of the main organisers of VoxSRC workshops.
\end{IEEEbiographynophoto}
\vspace{-35pt}
\begin{IEEEbiographynophoto}{Joon Son Chung}
is an assistant professor at Korea Advanced Institute of Science and Technology, where he is directing research in speech processing, computer vision and machine learning. He received the D.Phil. in Engineering Science from the University of Oxford. 
\end{IEEEbiographynophoto}
\vspace{-35pt}
\begin{IEEEbiographynophoto}{Arsha Nagrani} is a Senior Research Scientist at Google Research. She obtained her PhD from the VGG group in the University of Oxford, where her thesis won the ELLIS PhD Award. Her research focuses on cross-modal and multi-modal machine learning techniques for video recognition. Her work has been recognised by a Best Student Paper Award, Outstanding Paper Award, a Google PhD Fellowship and a Townsend Scholarship, and has been covered by major outlets such as The New Scientist, MIT Tech review and Verdict.
\end{IEEEbiographynophoto}
\vspace{-35pt}
\begin{IEEEbiographynophoto}{Andrew Brown} completed his PhD in computer vision and machine learning in the University of Oxford's Visual Geometry Group under the supervision of Professor Andrew Zisserman. 
His research during his PhD focused on end-to-end learning and audio-visual human-centric video understanding.
\end{IEEEbiographynophoto}
\vspace{-35pt}
\begin{IEEEbiographynophoto}{Jee-weon Jung} is a Postdoctoral Research Associate at Carnegie Mellon University, USA. He received his PhD degree from the University of Seoul, Republic of Korea. Before he joined Carnegie Mellon University, he was a research scientist at Naver Corporation, Republic of Korea. He has worked on speaker recognition, acoustic scene classification, audio spoofing detection, and other tasks. He was the main organiser of the Spoofing-Aware Speaker Verification Challenge, a special session at Interspeech 2022. He is one of the organisers of VoxSRC and ASVspoof since 2022.
\end{IEEEbiographynophoto}
\vspace{-35pt}
\begin{IEEEbiographynophoto}{Daniel Garcia-Romero} is a Principal Scientist at Amazon. Prior to that, he was a Senior Research Scientist at the Human Language Technology Center of Excellence (HLTCOE), Johns Hopkins University. He obtained his PhD in Electrical Engineering at the University of Maryland, College Park. His research interests are in the broad areas of speech processing, deep learning, and multi-modal person identification. For the past few years he has been working on deep neural networks for speaker, language recognition, and diarization. He is co-inventor of the x-vector embeddings that have set the state of the art in these fields.
\end{IEEEbiographynophoto}
\vspace{-35pt}
\begin{IEEEbiographynophoto}{Andrew Zisserman} is a Royal Society Research Professor and the Professor of Computer Vision Engineering at the Department of Engineering Science at the University of Oxford. His research interests have included multiple view geometry, audio and visual recognition, and large scale retrieval in images and video. His papers have won many best paper and test-of-time awards at international conferences.
\end{IEEEbiographynophoto}

\end{document}